\documentclass[a4paper,fleqn,usenatbib]{mnras}

\usepackage{newtxtext,newtxmath}
\usepackage[T1]{fontenc}
\usepackage{ae,aecompl}

\usepackage{graphicx}
\usepackage{amsmath}
\usepackage{amssymb}
\usepackage[usenames]{color}

\newcommand{\hh}{\mbox{\hspace{1.5mm}}}

\newcommand{\x}{\mbox{\hspace{3.0mm}}}

\newcommand{\e}{\mbox{\hspace{4.0mm}}}

\newcommand{\kms}{\mbox{km\,s$^{-1}$}}
\usepackage[usenames]{color}

\title[H$_2$O and SiO masers in OH\,231.8+4.2]{Time variations of H$_2$O and SiO masers in the proto-Planetary Nebula OH\,231.8+4.2}

\author[J. Kim et al.]{
Jaeheon Kim,$^{1}$\thanks{E-mail: jhkim@shao.ac.cn}
S.-H. Cho,$^{2,6}$
V. Bujarrabal,$^{3}$
H. Imai,$^{4,5}$
R. Dodson,$^{7}$
D.-H. Yoon$^{2}$
\newauthor 
and B. Zhang$^{1}$
\\
$^{1}$Shanghai Astronomical Observatory, Chinese Academy of Sciences, 80 Nandan rd, Shanghai 200030, China\\
$^{2}$Korea Astronomy and Space Science Institute, 776 Daedeok-daero, Daejeon 34055, Republic of Korea\\
$^{3}$Observatorio Astron\'{o}mico Nacional (OAN-IGN), Ap. 112, 28803 Alcal\'{a} de Henares, Spain\\
$^{4}$Center for General Education, Institute for Comprehensive Education, Kagoshima University, 1-21-30 Korimoto, Kagoshima 890-0065, Japan\\
$^{5}$Amanogawa Galaxy Astronomy Reserch Center, Graduate School of Science and Engineering, Kagoshima University, 1-21-35 Korimoto, Kagoshima 890-0065, Japan\\
$^{6}$Department of Astronomy, Yonsei University, 50 Yonsei-ro Seodaemun-gu, Seoul, 03722, Republic Of Korea\\
$^{7}$International Centre for Radio Astronomy Research, University of Western Australia, 35 Stirling Hwy, Australia}

\date{Accepted 2019 June 27. Received 2019 June 14; in original form 2018 December 25}

\pubyear{2019}

\begin{document}
\label{firstpage}
\pagerange{\pageref{firstpage}--\pageref{lastpage}}
\maketitle

\begin{abstract}

H$_2$O (22\,GHz) and SiO masers (43, 86, 129\,GHz) in the bipolar proto-planetary nebula OH\,231.8+4.2 were simultaneously monitored using the 21-m antennas of the Korean VLBI Network in 2009--2015. Both species exhibit periodic flux variations that correlate with the central star\rq{s} optical light curve, with a phase delay of up to 0.15 for the maser flux variations with respect to the optical light curve. The flux densities of SiO $v$ = 2, $J$ = 1$\rightarrow$0 and H$_2$O masers decrease with time, implying that they may disappear in 10--20 years. However, there seems to have been a transient episode of intense H$_2$O maser emission around 2010. We also found a systematic behaviour in the velocity profiles of these masers. The velocities of the H$_2$O maser components appear to be remarkably constant, suggesting ballistic motion for the bipolar outflow in this nebula. On the other hand, those of the SiO maser clumps show a systematic radial acceleration of the individual clumps, converging to the outflow velocity of the H$_2$O maser clumps. Measuring the full widths at zero power of the detected lines, we estimated the expansion velocities of the compact bipolar outflow traced by H$_2$O maser and SiO thermal line, and discussed the possibility of the expanding SiO maser region in the equatorial direction. All of our analyses support that the central host star of OH231.8 is close to the tip of the AGB phase, and that the mass-loss rate recently started to decrease because of incipient post-AGB evolution.

\end{abstract}



\section{Introduction}

OH\,231.8+4.2 (a.k.a. OH0739$-$14, IRAS~07399$-$1435, hereafter OH231.8) -- which has the common name of Calabash nebula, Rotten Egg nebula, or Juggler nebula -- is a particularly well-studied bipolar proto-planetary nebula (pPN) candidate. This object gives an extreme and puzzling conundrum, showing three varieties of stellar masers, namely, OH, H$_2$O and SiO, that usually occur in oxygen-rich asymptotic giant branch (AGB) stars, as well as various molecular lines of C-, S- and N-bearing species that appear in carbon-rich objects. Several authors have tried to resolve the identity of OH231.8 at diverse wavelengths over the past four decades, using information in X-ray, optical, near-infrared (NIR), mm-continuum, molecular line emissions and masers (e.g. X-ray; \citealt{1993ApJ...409..720T}, optical; \citealt{2004ApJ...616..519S} and references therein, IR \& mm-wave; \citealt{2018A&A...618A.164S} and references therein, masers; \citealt{2018MNRAS.476..520D} and references therein), and also giving some geometric models of the bipolar outflow \citep[][and references therein]{2017ApJ...843..108B}, since its discovery by \citet{1971ApL.....8...73T} who detected the OH emission at 1667\,MHz, which is seen in emission over a remarkably large velocity range of $\sim$\,100\,\kms. The OH profile is relatively flat in this range, i.e. an emission plateau.

\begin{figure*}
	\includegraphics[width=160mm]{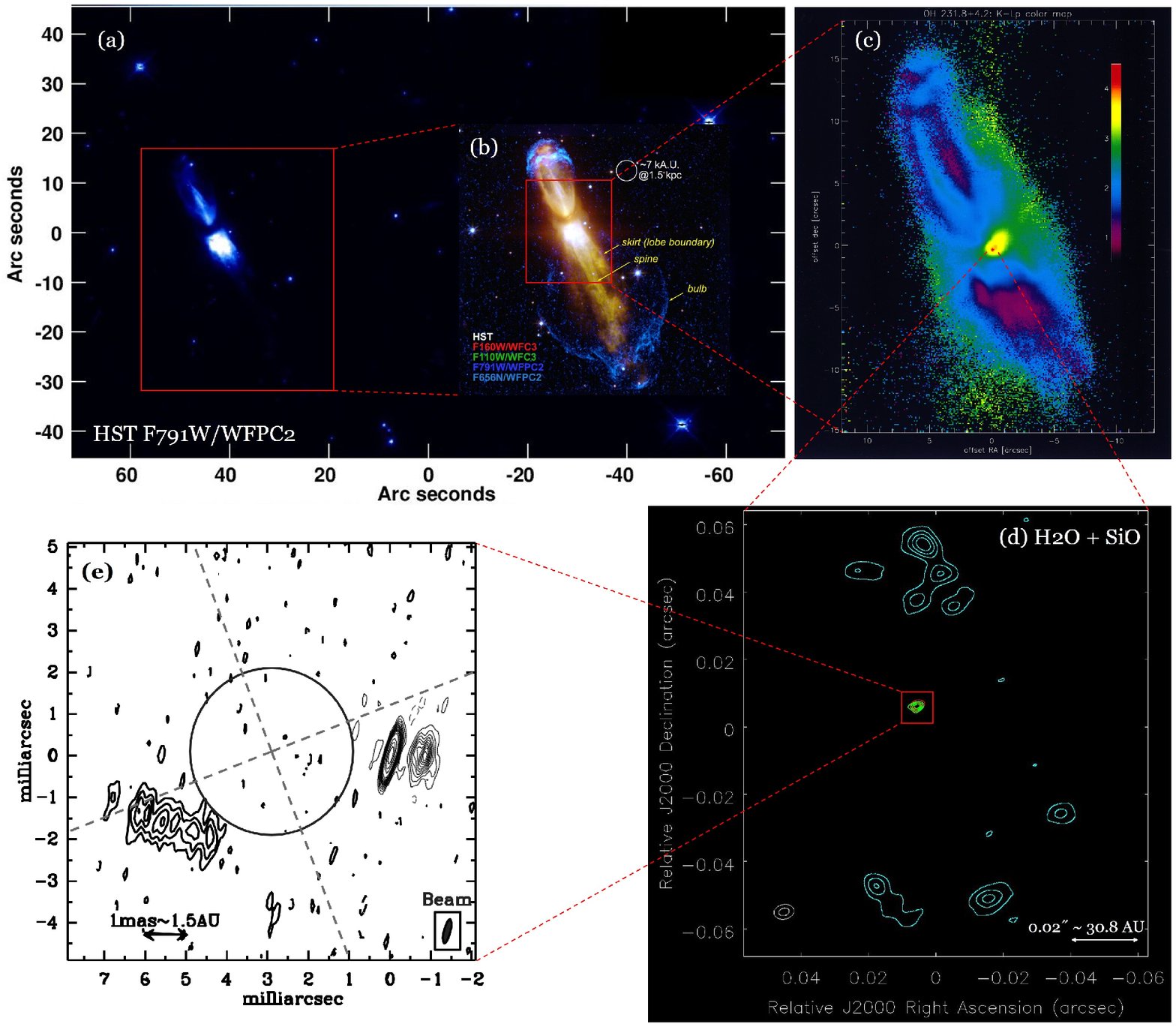}
    \caption{Images of OH231.8 in optical, NIR and radio wavelengths. (a): The continuum scattered light image in the F791W filter obtained from the Hubble Legacy Archive, showing a classic bipolar structure. (b): Overlaid $HST$ image of optical and NIR filters \citep[see][]{2017ApJ...843..108B}. (c): Close-up $K (2.2\,\micron)-L\rq{} (3.8\,\micron)$ color map \citep{1998AJ....116.1412K}, the reddest point in the map may be the position of the central star or near to it. (d) Close-up contour map of the central region obtained from the frequency phase referencing KVN observation of the H$_2$O, SiO $v$ = 1 and 2, $J$ = 1$\rightarrow$0 masers by \citet{2018MNRAS.476..520D}, the small red square indicates the astrometrically registered position of the SiO maser, and the cyan contour represents the H$_2$O maser features. (e): Spatial distribution of SiO $v$ = 2, $J$ = 1$\rightarrow$0 maser line obtained from the VLBA observation by \citet{2002AA...385L...1S}, the circle represents the angular extent of the central star, QX~Pup, and the dashed lines indicates the directions of the nebular symmetry axis and equator.}
    \label{fig:1}
\end{figure*}

Geometrically, OH231.8 has a fast expanding bipolar and bubble-like tenuous outflow seen in the H$\alpha$ image and a massive jet traced by the collimated CO molecular outflow, as the results of shocks propagating through the outer regions of the envelope. The central stellar system is deeply embedded in the dusty torus, and the bipolar outflow of this nebula flows out to north and south directions with immense deprojected velocities $\sim$\,430 (southern lobe) and $\sim$\,210 (northern lobe) \kms\ relative to the centre, traced by $^{12}$CO molecular line \citep{2001A&A...373..932A}. The majority of the gas flow observed today seems to stem from a sudden ejection event that happened only about 800 years ago. Fig.~\ref{fig:1} shows full and close-up images of OH231.8 in diverse wavelengths taken from the literature to demonstrate its bipolar geometry and show the central region. The nebula, in Figs.~\ref{fig:1}a and b, has an angular size of about 1 arcmin and is clearly elongated with axisymmetry at a position angle of $\sim$\,21$\degr$ \citep{2002A&A...389..271B,2017ApJ...843..108B}. The distance to OH231.8 is $\sim$\,1.54 kpc, derived from a trigonometric parallax measurement of the H$_2$O masers by multi-epoch Very Long Baseline Array (VLBA) observations \citep{2012IAUS..287..407C}. In addition, the presence of a Mira variable (M9III--M10III, named QX~Pup) and the possibility of the existence of a compact companion have been suggested by several authors \citep{1981PASP...93..288C,1985ApJ...297..702C,1992ApJ...398..552K,1998AJ....116.1412K}, and direct characterization of the companion as an A0V star was reported by \citet{2004ApJ...616..519S,2006ApJ...646L.123M}.

The spatial distributions of the H$_2$O and SiO masers have been imaged only a few times at sub-arcsecond scales with the VLBA \citep[e.g.][]{2002AA...385L...1S,2007AA...468..189D,2012IAUS..287..407C,2012AA...540A..42L}, focusing on the inner region of OH231.8. \citet{2018MNRAS.476..520D}, recently, presented the astrometrically registered map of H$_2$O (north and south) and SiO $v$ = 1 and 2, $J$ = 1$\rightarrow$0 (in centre) masers for this nebula using the Korean VLBI network (KVN), although its angular resolution is insufficient to interpret the disk structure traced by the SiO masers (Fig.~\ref{fig:1}d). This map, interestingly, resembles the view of the optical + NIR image (Fig.~\ref{fig:1}b) zoomed in by a factor of $\sim$\,500. The H$_2$O maser clumps are distributed with several compact features mainly in two regions to the north and south directions. These fall at $\sim$\,50 mas ($\sim$\,77 AU) from the central region of the nebula emitted SiO masers, and trace well a compact bipolar outflow. SiO masers are located in the middle of both H$_2$O maser emitting regions, identifying the position of the Mira variable QX~Pup, expected to be the primary star of the binary system and at the centre of the nebula. In particular, \citet{2002AA...385L...1S} suggest that the SiO maser features of the central star QX~Pup forms an equatorial torus, with the structure elongated in the direction perpendicular to the symmetry axis of the nebula, and is undergoing the rotating + infalling motion, based on the VLBA image of SiO $v$ = 2, $J$ = 1$\rightarrow$0 maser emission as shown in Fig.~\ref{fig:1}e.

In general, it is very unusual that both H$_2$O and SiO masers (especially SiO), that mainly occur in oxygen-rich envelopes during the AGB stage, are detected in binary/symbiotic systems with jet-like bipolar outflows in a brief pPN stage. H$_2$O molecules are expected to be photodissociated to OH by the UV photons and the velocity coherence of the circumstellar atmosphere of the primary which generates SiO masers may be disrupted by the tidal interaction between the primary and the secondary. However, they are still found in the circumstellar conditions of OH231.8. In terms of the evolution, it is also very interesting that this pPN-like nebula/outflow has not a post-AGB star at the centre, but an AGB star, i.e. a Mira variable that is still experiencing active mass ejection, in the central part of  the nebula. \citet{2004ApJ...616..519S} proposed that most circumstellar material in OH231.8 is the result of mass-loss from the central primary star, QX~Pup, whose slow, but dense, AGB wind has been partially trapped by its compact companion, forming an accretion disk, and re-ejected by the latter as a collimated, fast ($\gtrsim$ 400 \kms) wind in the direction perpendicular to the orbital plane. Therefore, we concentrate on the masers around the primary star, believed to be the Mira variable, of the central binary system of OH231.8 to understand the origin of this nebula and give some more information on how it developed. The secondary A0V companion cannot be directly observed, presumably due to a dense circumstellar accretion disk obscured along our line of sight. 

Furthermore, one of the important viewpoints for considering the origin of OH231.8 is how the physical properties of the molecular atmosphere of OH231.8 vary in time. The intensity and velocity variations of both H$_2$O and SiO masers can be a useful probe for the circumstellar atmosphere of OH231.8, since the SiO masers arise from the turbulent region between stellar photosphere and dust condensation zone \citep[$\sim$\,2--4 $R_{\ast}$][]{1994ApJ...430L..61D} and the H$_2$O masers are located in the inner parts of the circumstellar envelope (CSE) and above the dust forming layer between 10 and 20 $R_{\ast}$ \citep{1996ApJS..106..579B}. Therefore, it is important to perform simultaneous observations of H$_2$O and SiO masers together because the time variations of intensities and line profiles in the H$_2$O and SiO masers imply their tight correlation with the stellar pulsation \citep[see][for details]{2014AJ....147...22K}. In addition, the time variations of their radial velocities allow us to infer the kinematics of the circumstellar atmosphere.

In this paper, we present the results of simultaneous monitoring observations of H$_2$O and SiO masers toward OH231.8 for about 7 years from 2009, to understand the characteristics of the H$_2$O and SiO masers and their relationship to the bipolar outflow including the inner region of OH231.8. The observations of OH231.8 are described in Section 2 and their results are presented in Section 3. In Section 4, we analyse the periodic variations of peak/integrated flux densities and peak/mean velocities of the H$_2$O and SiO masers based on our detections of a periodic variability in these maser lines. Conclusions are summarized in Section 5.

\section{Observations}

Simultaneous monitoring and survey observations were carried out in H$_2$O 6$_{1,6}$--5$_{2,3}$ (22\,GHz, K band), $^{28}$SiO $v$ = 1, 2, 3, 4, $J$ = 1$\rightarrow$0 (43\,GHz, Q band), $^{28}$SiO $v$ = 1, 2, $J$ = 2$\rightarrow$1 (86\,GHz, W band), $^{28}$SiO $v$ = 1, 2, $J$ = 3$\rightarrow$2 (129\,GHz, D band) maser lines, $^{28}$SiO $v$ = 0, $J$ = 1$\rightarrow$0, 2$\rightarrow$1, 3$\rightarrow$2 thermal lines and the isotopologue $^{29}$SiO $v$ = 0, $J$ = 1$\rightarrow$0 line, with the Yonsei, Ulsan and Tamna 21-m radio telescopes of the KVN from 2009 June to  2015 December. The list of the observed H$_2$O and SiO transitions and their frequencies are presented in Table~\ref{tab:1}. Monitoring observations were conducted in the same observational mode as used for the post-AGB stars as reported previously \citep[see][]{2016JKAS...49..261K}.

\begin{table}
	\centering
	\caption{Observed molecular transitions and their rest frequencies}
	\label{tab:1}
	\begin{tabular}{ccr}
		\hline
		Molecule      & Transition                 & Frequency (GHz) \\
		\hline
		H$_2$O        & 6$_{1,6}$--5$_{2,3}$       &  22.235080\e \\
		$^{28}$SiO\hh & $v$=0, $J$=1$\rightarrow$0 &  43.423858\e \\
					  & $v$=1, $J$=1$\rightarrow$0 &  43.122080\e \\
					  & $v$=2, $J$=1$\rightarrow$0 &  42.820587\e \\
					  & $v$=3, $J$=1$\rightarrow$0 &  42.519379\e \\
					  & $v$=4, $J$=1$\rightarrow$0 &  42.218456\e \\
					  & $v$=0, $J$=2$\rightarrow$1 &  86.846998\e \\
					  & $v$=1, $J$=2$\rightarrow$1 &  86.243442\e \\
					  & $v$=2, $J$=2$\rightarrow$1 &  85.640452\e \\
					  & $v$=0, $J$=3$\rightarrow$2 & 130.268706\e \\
					  & $v$=1, $J$=3$\rightarrow$2 & 129.363359\e \\
					  & $v$=2, $J$=3$\rightarrow$2 & 128.458891\e \\
		$^{29}$SiO\hh & $v$=0, $J$=1$\rightarrow$0 &  42.879916\e \\
		\hline
	\end{tabular}
\end{table}

Only left circular polarized K/Q band receivers were available during the observation periods of 2009 June -- 2012 May. Since 2012 June, when W/D band receivers were installed, the cryogenic K, Q, W bands High Electron Mobility Transistor (HEMT) receivers and D band Superconductor-Insulator-Superconductor (SIS) receiver were used. These have both right and left circularly polarized feeds \citep{2013PASP..125..539H}. For all the observations, however, we used only the left circular polarized feed to get uniform data. The K, Q, W and D bands receivers yielded system noise temperatures of 90--330\,K (K), 130--380\,K (Q), 230--330\,K (W) and 220--580\,K (D), respectively. Detailed information for the half power beam widths (HPBW) and aperture efficiencies of three KVN antennas at four bands can be found at the KVN website\footnote{\url{https://radio.kasi.re.kr/kvn/status_report_2018/home.html}}. Here we present just the averaged values of three antennas, i.e. the averaged HPBW and the aperture efficiencies are: 123\arcsec/0.58 (K), 62\arcsec/0.61 (Q), 32\arcsec/0.50 (W) and 23\arcsec/0.35 (D), for the four frequencies respectively. The pointing and tracking were found to be accurate to $\sim4\arcsec$ and were checked about every 2--3 hrs by observing the nearby strong SiO maser source, the supergiant VY CMa.

For spectral analyses, we originally used a 4096-channel digital filter bank (DFB) with chosen total bandwidths of 4 $\times$ 64\,MHz for the K and Q bands (H$_2$O, $^{28}$SiO $v$ = 1, 2, and $^{29}$SiO $v$ = 0, $J$ = 1$\rightarrow$0 lines), producing the velocity resolutions of 0.21\,\kms (K) and 0.11\,\kms (Q), until 2012 May. After upgrading to the simultaneous four band system on 2012 June, we have utilized 4 $\times$ 32\,MHz for the K/Q bands (H$_2$O and $^{28}$SiO $J$ = 1$\rightarrow$0 lines) and 2 $\times$ 64\,MHz for the W and D bands ($^{28}$SiO $J$ = 2$\rightarrow$1 and $^{28}$SiO $J$ = 3$\rightarrow$2 lines) with the velocity resolutions of 0.11\,\kms (K) and 0.05\,\kms (Q), 0.05\,\kms (W) and 0.036\,\kms (D). All the spectra were Hanning-smoothed to a velocity resolution of 0.42--0.44\,\kms\ to improve the signal-to-noise ratio and provide a uniform velocity distribution between different bands. The interval between two consecutive observing sessions was roughly 2--5 months depending on the weather conditions in an assigned date and the maintenance schedule.

The calibration was carried out by the chopper wheel method, to correct the atmospheric attenuation and the antenna gain variations, and then the sky dipping curve analysis was conducted to correct the atmospheric opacity toward the target direction, to yield an antenna temperature $T_{\rm A}^{\ast}$. The total integration time for each spectrum was 90--120 minutes (ON + OFF) to achieve the 1-$\sigma$ sensitivity of 0.01--0.05\,K. The averaged conversion factors of the three KVN telescopes from an antenna temperature to a flux density are about 13.8\, Jy\,K$^{-1}$ (K), 13.1\,Jy\,K$^{-1}$ (Q), 15.9\,Jy\,K$^{-1}$ (W), and 22.8\,Jy\,K$^{-1}$ (D), respectively. In Column 7 of Tables~\ref{tab:a1} and \ref{tab:a2}, we present conversion factors that actually applied in each observation (see the following Section).

\section{Observational results}

We obtained spectra of OH231.8 at all 13 transitions of H$_2$O and SiO masers during the period of 2009 June -- 2015 December. Figs.~\ref{fig:b1}--\ref{fig:b5} show the line profiles of all detected transitions at all the epochs during the monitoring. The observed dates (YYMMDD) and maser transitions are indicated in each spectrum. The systemic radial velocity ($\sim$\,35\,\kms) was determined from molecular line ALMA observation \citep{2018A&A...618A.164S}. Relevant parameters for each spectrum are listed in Tables~\ref{tab:a1}--\ref{tab:a3}. As exhibited in these figures and tables, H$_2$O, SiO $v$ = 0, 1, 2, $J$ = 1$\rightarrow$0 and SiO $v$ = 0, 1, $J$ = 2$\rightarrow$1 lines are detected at 2--24 epochs, depending on transitions. Optical light curve (OLC) data provided by the American Association of Variable Star Observers (AAVSO) were used, to investigate the correlation of H$_2$O and SiO masers with the optical stellar emission. The OLC shows a quite regular begavior, with a period of about 550 days and a maximum amplitude of 12.1--15.0$^m$. Our monitoring extends over about four periods, Details of the optical variability are reported in Sect. 4.1.

\begin{figure}
    \includegraphics[width=\columnwidth]{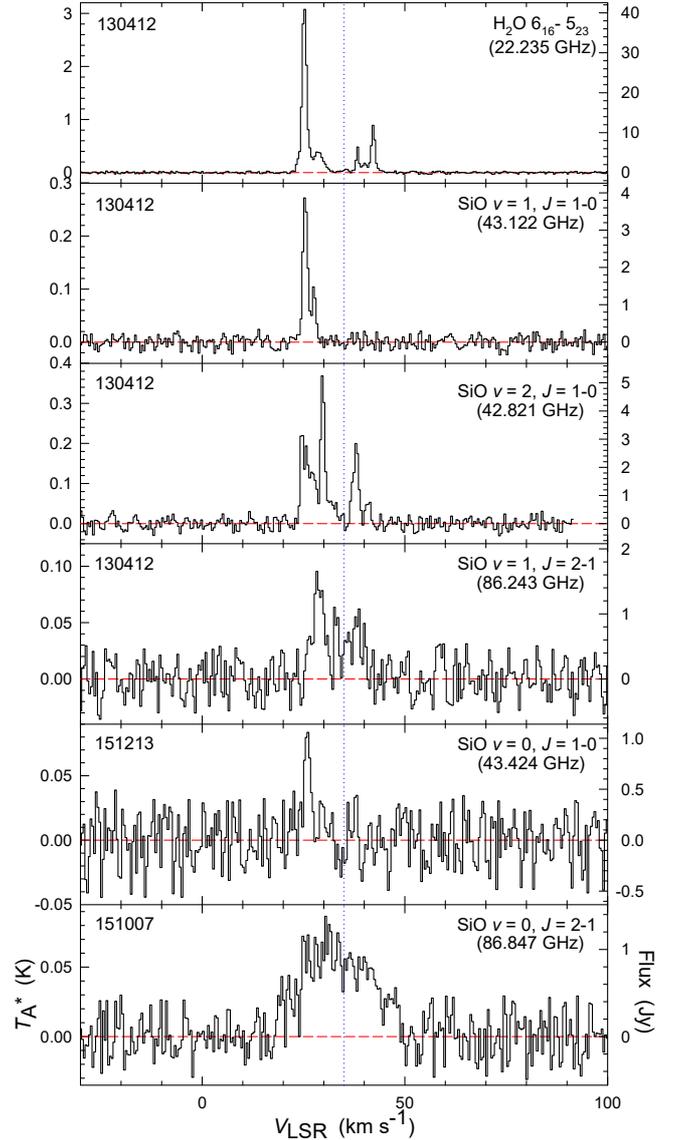}
    \vspace{-5mm}
    \caption{Sample spectra of the detected lines. The intensity is given in units of the antenna temperature $T_{\rm A}^{\ast}$ (K), and the abscissa is $V_{\rm LSR}$ (\kms). The vertical blue dotted lines mark the systemic radial velocity.}
    \label{fig:2}
\end{figure}

Fig.~\ref{fig:2} shows a set of sample spectra of detected lines. Both H$_2$O, SiO $v$ = 1, 2, $J$ = 1$\rightarrow$0 and SiO $v$ = 1, $J$ = 2$\rightarrow$1 maser lines were simultaneously obtained on 2013 April 12 (130412) very close to the optical maximum ($\phi$ = 2.98, the optical maximum phase $\phi$ = 0.0, 1.0, 2.0, 3.0, 4.0, 5.0), and SiO $v$ = 0, $J$ = 1$\rightarrow$0 and $J$ = 2$\rightarrow$1 lines were obtained on 2015 December 13 (151213, $\phi$ = 4.74) and 2015 October 7 (151007, $\phi$ = 4.62). 

The H$_2$O maser spectra commonly exhibit four peaked line profiles at $V_{\rm LSR}$ of 25.4, 29.1, 38.3 and 42.3\,\kms, distributed in two main groups with the blue- and redshifted features relative to the systemic velocity of the object. The peak intensity of the blueshifted component at 25.4\,\kms\ is almost five times stronger than that of the red components, and their emission strength varies with time by a factor of 5--8. This feature was originally detected by \citet{1976ApJ...204..415M} and the most redshifted feature at 42.3\,\kms\ was first presented by \citet{1977AAS...30..145G}. Both of the two features continue to be seen up to the current date (see Fig.~\ref{fig:b1}). The weak components at 29.1 and 38.3\,\kms\ have been recently presented by the VLBI observations of \citet{2007AA...468..189D,2012IAUS..287..407C,2018MNRAS.476..520D}. 

\begin{figure*}
	\includegraphics[width=140mm]{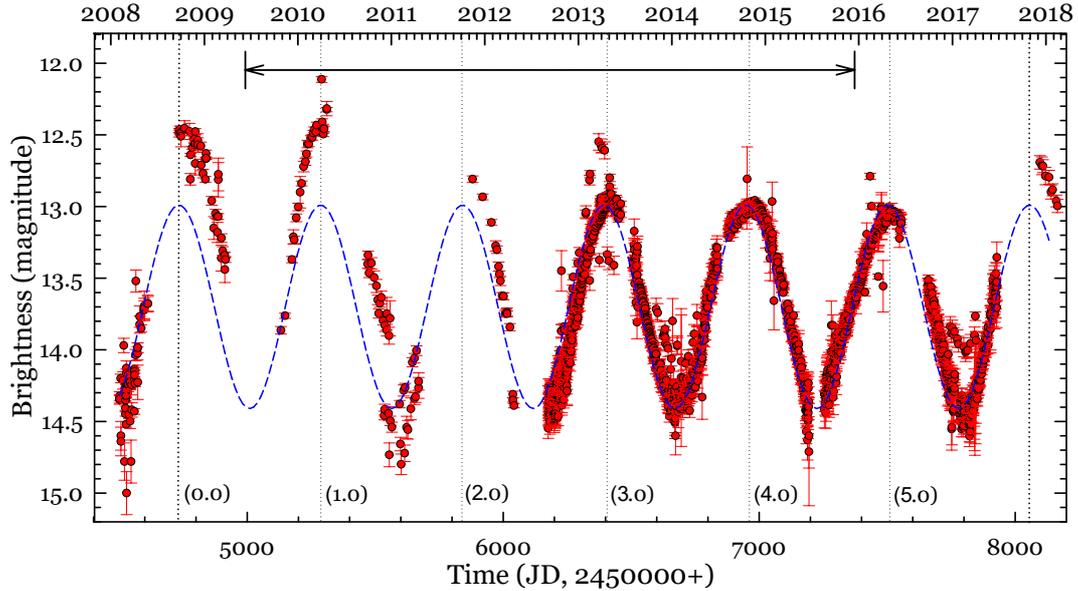}
    \caption{The OLC of the central star QX~Pup in OH231.8 nebula from AAVSO (red circles), showing a period of 554 $\pm$ 9 days. The optical maximum phases ($\phi$) are represented by vertical dotted lines and the numbers in parentheses. The horizontal arrow line indicates our monitoring period from 2009 June 07 to 2015 December 13. The corresponding best fitting sinusoidal wave curve is shown as blue dashed line.}
    \label{fig:3}
\end{figure*}

SiO $v$ = 1, $J$ = 1$\rightarrow$0 maser emission is detected at an averaged peak flux density of 1.72\,Jy at $V_{\rm LSR}$ = $\sim$\,25.1\,\kms, at only eight epochs during mainly 2013--2014 (see Fig.~\ref{fig:b2}). The averaged peak velocity of 25.1\,\kms\ is very close to the velocity of the blueshifted H$_2$O maser peak (i.e. not the midpoint of the H$_2$O line profile). This velocity distribution is very unusual because, in general, the peak velocity of SiO maser emission is centred on the midpoint of the 22 GHz H$_2$O maser profile in most AGB stars \citep[][and references therein]{2014AJ....147...22K}. There is a sudden strong emission around 2013 April 12. This maser was not detected in data taken up till and including mid-2012 and was only detected from 2012 October. It showed the strongest emission on 2013 April 12, and then rapidly weakened and disappeared. At that time, the peak flux density of the SiO $v$ = 1, $J$ = 1$\rightarrow$0 maser shows 3.87\,Jy at $V_{\rm LSR}$ = 25.5\,\kms\ comparable to the intensity of SiO $v$ = 2, $J$ = 1$\rightarrow$0 maser. Apart from the tentative detections by \citet{1994PASJ...46..629T,1998AAS..127..185N,2002AA...385L...1S,2007AA...468..189D,2010ApJS..188..209K}, this emission spike was indubitably detected only once by the observation on 1983 December of \citet{1991AA...242..211J}, when it appeared four times stronger than our result. This sudden maser brightening may be due to short-lived events of enhanced mass-loss from the central Mira variable in relation to the stellar pulsation. Another possible interpretation is the maser powering effect due to the orbital motion of its compact companion, passing through the periastron. These possibilities are described in the following section.

SiO $v$ = 2, $J$ = 1$\rightarrow$0 maser was detected relatively frequently in 21 epochs, showing 2--4 peaked profiles with a strong flux variation by a factor of 4--5 (see Fig.~\ref{fig:b3}). Parameters of the strongest double-peaked profile only are listed in Table~\ref{tab:a1}. Such a SiO maser profile with 2--4 peaks is rarely detected for an evolved star, while some supergiants and binary/symbiotic systems often show these type of line profiles, implying rotation of the SiO maser shell (e.g. VX~Sgr; \citealt{1998ApJ...494..400D}, NML~Cyg; \citealt{2000ApJ...545L.149B}, R~Aqr; \citealt{2000ApJ...543L..81H}, IK~Tau; \citealt{2005ApJ...625..978B}). The full line width of SiO $v$ = 2 maser components in OH231.8 increases with time, and new emission lines inside double-peaked features arise and vanish. 

For the SiO $v$ = 1, $J$ = 2$\rightarrow$1 maser (Fig.~\ref{fig:b4}), the lines were detected at the early six epochs out of a total of 13 epochs, and have not been detected since the observations of 2013 November 27 (131127). The detected lines exhibit double-peaked profile groups that are separated with the blue- and redshifted components like the $v$ = 2, $J$ = 1$\rightarrow$0 profiles. The date when the maximum flux ($\sim$\,1.65\,Jy) of the SiO $v$ = 1, $J$ = 2$\rightarrow$1 maser was observed was the same as the date (2013 April 12) when the maximum fluxes of SiO $v$ = 1 and 2 $J$ = 1$\rightarrow$0 masers were observed. On the other hand, the H$_2$O maser appears to be insensitive to the presence of strong SiO maser intensity, showing only a gradual decrease with the periodic maser flux variation relative to the OLC. These facts corroborate that SiO and H$_2$O masers of OH231.8 are distributed at different locations with respect to the centre of the nebula in agreement with the VLBI observational result of \citet{2018MNRAS.476..520D}.

Meanwhile, SiO $v$ = 0, $J$ = 1$\rightarrow$0 emission was detected for the first time in four different epochs (Fig.~\ref{fig:b5}). This emission line exhibits weak, narrow features at a similar velocity range with SiO $v$ = 1, $J$ = 1$\rightarrow$0 line. SiO $v$ = 0, $J$ = 2$\rightarrow$1 line is also detected in two out of three epochs, showing the signature of thermal circumstellar emission, e.g., broad and parabolic line profiles. Detection of the SiO $v$ = 0, $J$ = 2$\rightarrow$1 has been previously published by \citet{1987ApJ...321..888M}. For all detected maser transitions, the flux and velocity variations with time are discussed in Section 4.

\section{Analyses and discussion}
\subsection{The optical variability}

The primary star of the central binary system of OH231.8 nebula is classified as a Mira-type long period variable star. Previous photometric studies of the nebula conclude that the pulsation period is $\sim$\,661 days in $JHKL$ infrared observations (\citealt{1983MNRAS.203.1207F}, re-calculated by \citealt{1984ApJ...276..646B}), $\sim$\,684 days in 1667\,MHz OH monitoring \citep{1984ApJ...276..646B}, $\sim$\,692 days and $\sim$\,708 days in $2.2\,\micron$ monitoring \citep{1987Natur.325..787R,1992ApJ...398..552K}. From these results over the last 25--30 years, the pulsation period could be estimated to be 690 days. Recently, AAVSO has been providing the optical data for the central star QX~Pup from 2008 February up to the present. The complementary data were taken through the Cousins $I$ filter. The magnitude uncertainty of the data is on average less than 0.1. Fig.~\ref{fig:3} shows its OLC and the dynamic fitting curve with simple sinusoidal waveforms over a period of about 10 years (2008--2018). From the fitting, we derive a period of 554 $\pm$ 9 days, giving the following ephemeris: 
\begin{equation}
\rm Max. = JD\,2\,450\,000 + 4\,740, 5\,288, 5\,839, 6\,406, 6\,962, 7\,510, 8\,059.
\end{equation}

According to Fig.~\ref{fig:3} and the above maximum values (eq. 1), we could estimate that the period of the last decade is almost 140 days shorter than those of 25--30 years ago. The shortening of the pulse cycle over time is related with the mass-loss rate of the central star. During the final phase for AGBs, the star has a thermally unstable helium-burning shell, punctuated by briefly fierce shell flashes \citep{1981ApJ...247..247W}. This gives rapid changes in surface luminosity of the star, and during each pulse, most of the materials are ejected from the star, giving an empirical formula relating the mass-loss rate to the pulsation period \citep{1993ApJ...413..641V}. We adopt each maximum value (eq. 1) of the OLC in our subsequent analyses of H$_2$O and SiO maser properties to investigate the correlation of both masers with optical phases.

\subsection{Time variations of the maser fluxes}

We mainly focus on the time monitoring results of the detected H$_2$O, SiO $v$ = 1, 2, $J$ = 1$\rightarrow$0 and SiO $v$ = 1, $J$ = 2$\rightarrow$1 masers. 

\begin{figure}
    \includegraphics[width=\columnwidth]{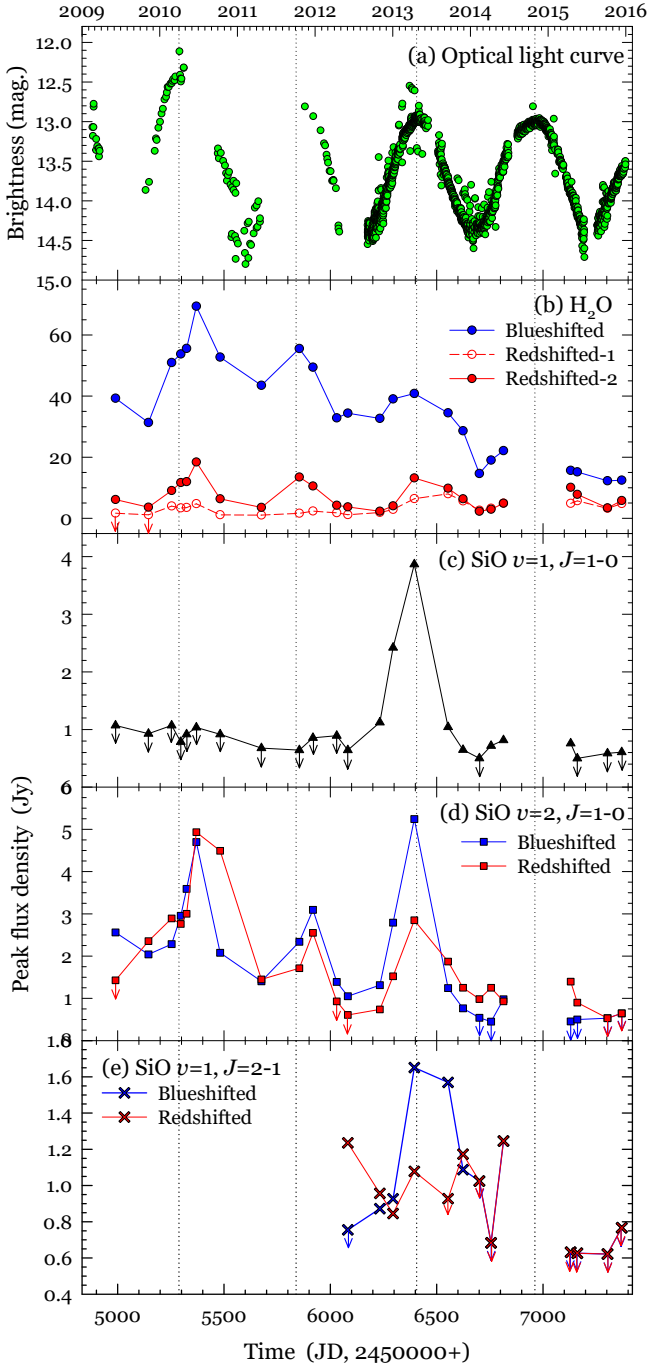}
    \caption{Peak flux densities of H$_2$O, SiO $v$ = 1, 2, $J$ = 1$\rightarrow$0 and SiO $v$ = 1, $J$ = 2$\rightarrow$1 masers of OH231.8 during 2009--2016. (a): Green circles show the optical magnitude that have been obtained from the AAVSO. The optical maxima are indicated by the vertical dotted lines. (b)--(e): The symbols used are described in the legend inside the plots. The downward arrows indicate the upper limit (3$\sigma$) of undetected epochs.}
    \label{fig:4}
\end{figure}

\begin{figure}
    \includegraphics[width=\columnwidth]{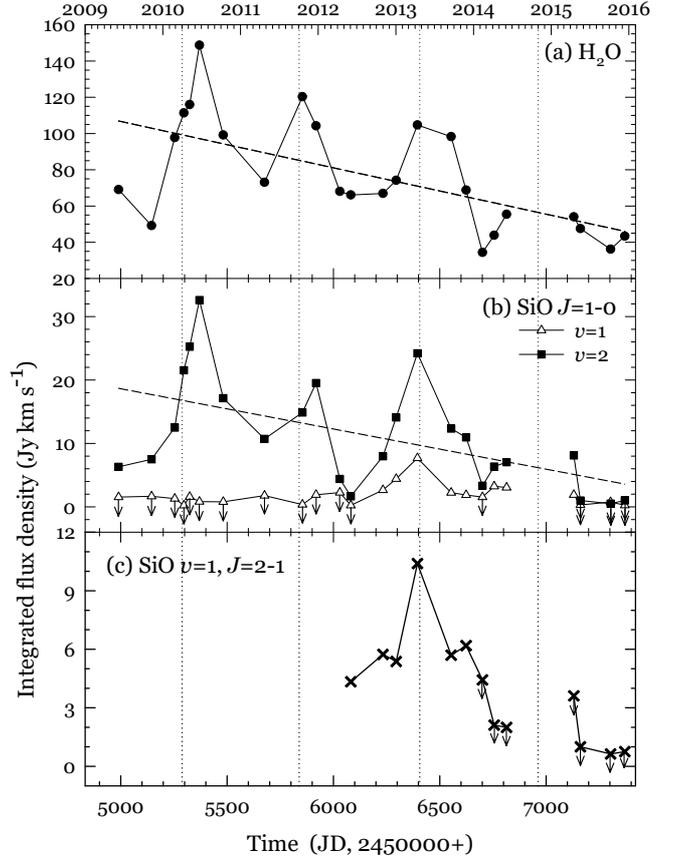}
    \caption{Integrated flux densities of H$_2$O, SiO $v$ = 1, 2, $J$ = 1$\rightarrow$0 and SiO $v$ = 1, $J$ = 2$\rightarrow$1 masers of OH231.8 during 2009--2016. (a) The dashed line represents the fitting result of a linear regression by least-squares approximation. (b) The symbols used are described in the legend inside the plots. The dashed line represents the fitting result of a linear regression for the SiO $v$ = 2, $J$ = 1$\rightarrow$0 maser. The downward arrows indicate the integrated values that are applied as upper limits (3$\sigma$) fluxes of undetected epochs.}
    \label{fig:5}
\end{figure}

Figs.~\ref{fig:4} and ~\ref{fig:5} show periodic variations of the peak and integrated flux densities of H$_2$O, SiO $v$ = 1, 2, $J$ = 1$\rightarrow$0 and SiO $v$ = 1, $J$ = 2$\rightarrow$1 maser lines with the OLC. There was an interruption of approximately one year in 2014--2015 during the monitoring due to the system upgrade and start-up of the KVN Key Science Projects. Nevertheless, we could find that periodic flux variations of H$_2$O and SiO $v$ = 2, $J$ = 1$\rightarrow$0 masers follow the OLC of the central star well, although they are by far less evident than that of the OLC. Both maser fluxes show a long-term trend that decreases over time. However, in the case of SiO $v$ = 1, $J$ = 1$\rightarrow$0, there is no evidence for periodic variability. It showed only one strong flux peak near the optical maximum phase (MJD 6406, the modified Julian date; JD-2450000). In the case of SiO $v$ = 1, $J$ = 2$\rightarrow$1 maser, the number of observations is relatively small and the detection rate (50 \%) is too low to discuss any periodic pattern. However, the highest flux value of the blueshifted peak of this maser emission was also detected near MJD 6406, which had the optical maximum phase, as in the other two SiO masers.

To check these long-term trends, we also investigated previous observations. In addition, further KVN single dish observation was made on 2018 September 8 (= MJD 8370, $\phi$ = 5.57) to determine how these trends were reflected at the most recent date. Table~\ref{tab:2} gives the peak flux densities and their velocities of the H$_2$O and SiO $v$ = 1, 2, $J$ = 1$\rightarrow$0 masers for OH231.8 from publications and our recent observation. These results are displayed in Fig.~\ref{fig:6} with our observed data. 

\begin{table}
	\centering
	\caption{Published observations of the H$_2$O and SiO masers in OH231.8}
	\label{tab:2}
	\begin{tabular}{cccc}
	\hline
Date & $S_{\rm peak}$ & $V_{\rm peak}$ & Reference \\
     & (Jy)           & (\kms)         &           \\
	\hline
    \multicolumn{4}{c}{H$_2$O} \\
    \hline
1975 May  & 8.1        & 25.4       & \citet{1976ApJ...204..415M} \\
1977 Feb. & 9.0, 4.5   & 25.0, 42.0 & \citet{1977AAS...30..145G} \\
1980 Jan. &	3.1        & 25.0       & \citet{1984ApJ...276..646B} \\ 
1984 Api. & 16.1, 5.8  & 24.1, 45.3 & \citet{1988AA...191..283E} \\
1991 Oct. & 4.2        & 22.1       & \citet{2001PASJ...53..517T} \\
1992 May  & 13.0, 3.1  & 23.1, 44.5 & \citet{2001ApJ...557L.109G} \\
2002 Nov. & 12.3, 2.9  & 25.5, 42.9 & \citet{2007AA...468..189D} \\
2005 Dec. & 24.0, 10.0 & 26.3, 41.9 & \citet{2008PASJ...60.1077S} \\
2009 Mar. & 19.2, 6.8  & 25.6, 41.6 & \citet{2012AA...540A..42L} \\
2009 Jun. & 39.3, 6.1  & 25.7, 42.1 & \citet{2010ApJS..188..209K} \\
2010 May  & 28.0, 8.0  & 25.4, 41.9 & \citet{2012IAUS..287..407C} \\
2017 Jan. & 1.4, 2.0   & 28.4, 41.8 & \citet{2018MNRAS.476..520D} \\
2018 Sep. & 3.1, 2.2   & 28.4, 40.5 & J. Kim (in private, 2018) \\
	\hline
	\multicolumn{4}{c}{SiO $v$ = 1, $J$ = 1$\rightarrow$0} \\
    \hline
1983 Dec. & 13.8     & 25.8     & \citet{1991AA...242..211J} \\
1991 Api. & $<8.0$   & $\cdots$ & \citet{1994PASJ...46..629T} \\
1993 Sep. & $<5.1$   & $\cdots$ & \citet{1998AAS..127..185N} \\
2000 May  & $<0.1$   & $\cdots$ & \citet{2002AA...385L...1S} \\
2009 Jun. & $<1.1$   & $\cdots$ & \citet{2010ApJS..188..209K} \\
2017 Jan. & 0.6      & 35.8     & \citet{2018MNRAS.476..520D} \\
2018 Sep. & $<0.7$   & $\cdots$ & J. Kim (in private, 2018) \\
	\hline
	\multicolumn{4}{c}{SiO $v$ = 2, $J$ = 1$\rightarrow$0} \\
    \hline
1993 Sep. & 8.2, 6.6  & 27.2, 34.1 & \citet{1998AAS..127..185N} \\
2000 May  & 19.0, 7.0 & 30.0, 41.0 & \citet{2002AA...385L...1S} \\
2003 Jul. & 0.1       & 27.0       & \citet{2007AA...468..189D} \\
2009 Jun. & 2.6       & 31.2       & \citet{2010ApJS..188..209K} \\
2017 Jan. & 0.2, 0.8  & 27.6, 35.4 & \citet{2018MNRAS.476..520D} \\
2018 Sep. & 1.2, 1.4  & 28.1, 34.0 & J. Kim (in private, 2018) \\
	\hline
	\end{tabular}
\end{table}

\subsubsection{H$_2$O maser}

During monitoring periods, we obtained 24 H$_2$O maser spectra, predominantly showing three  or sometimes four features, one or two blue- and the other two redshifted with respect to the stellar velocity (Figs.~\ref{fig:2} and \ref{fig:b1}). The blueshifted feature is a blend of two individual maser lines. The weak blueshifted line close to the stellar velocity is difficult to distinguish clearly because it is sometimes embedded in the wing of a strong line feature or not appeared. Therefore, this line is excluded from the discussion.

In Fig.~\ref{fig:4}b, it is presented the peak flux variation of three-peaked H$_2$O maser components. We found that the periodic flux variations of the H$_2$O maser correlate well with the OLC. The maximum peak fluxes of the H$_2$O maser are represented at the optical phase (expressed in cycles) $\phi$ = 1.15, 2.03, 2.98, i.e. the normalized phase lag of 0--0.15. It is clearly apparent that H$_2$O maser emission of OH231.8 is closely related to the stellar pulsation of the central primary star with a small phase lag, even though there are no data around the fourth optical maximum between 2014 June 5 (= MJD 6814) and 2015 April 17 (= MJD 7130). This may be because H$_2$O molecules are mainly pumped by collisions with particles of the circumstellar gas in the outer region of the dust shell and the propagation of shock waves generated by the stellar pulsation through this region can enhance maser excitation. An alternative explanation was proposed by \citet{2012A&A...546A..16R}. They suggest that a periodic H$_2$O maser flux variation is affected by changing the stellar IR field so as to control the maser pumping efficiency. This can explain a zero or short phase lag in the maser flux variation. From a result of our monitoring observations only, the scenario that H$_2$O maser excitation changes are closely related to the stellar IR variability seems more reliable. However, we can not conclude on which scenario is correct, because our monitoring results only present changes over a limited period of time (three periodic cycle in about seven years). A longer monitoring will be needed.

From previous statistical studies \citep[][and references therein]{2014AJ....147...22K} and long-term monitoring observations of individual stars \citep[][and references therein]{2005A&A...437..127L}, clear correlations were presented between the flux density of H$_2$O maser emission and the optical phase of the central Mira variable, typically showing a phase lag of about 0.01--0.4. Thus, the performance of the H$_2$O maser in OH231.8 resembles that around Mira variable, although it is not clear. Therefore, one can consider that this maser is still associated with a CSE of the central Mira variable rather than the fast bipolar outflow that is visible in the optical observations and driven by the companion star in the binary system. 

Furthermore, as shown in Figs.~\ref{fig:4}b and ~\ref{fig:5}a, the peak and integrated intensities of the H$_2$O maser present a gradual decrease along with the decline of optical amplitude with time. This decreasing pattern can be expressed by simple linear regression analysis as follows (Fig.~\ref{fig:5}a);
\begin{equation}
f$(MJD)$ = 234.622 - (2.559 \times 10^{-2})\,$MJD$,
\end{equation}
where MJD is the modified Julian date (JD $-$2450000) and $f$(MJD) is the integrated flux density (Jy km s$^{-1}$). If this pattern of H$_2$O maser emission is simply continued, it will finally disappear or become weaker than $\sim$\,1\,Jy\,\kms\ around 2020--2021. The H$_2$O maser fluxes detected by \citet{2018MNRAS.476..520D} and J. Kim (in private, 2018) also follow this pattern. 

\begin{figure*}
	\includegraphics[width=160mm]{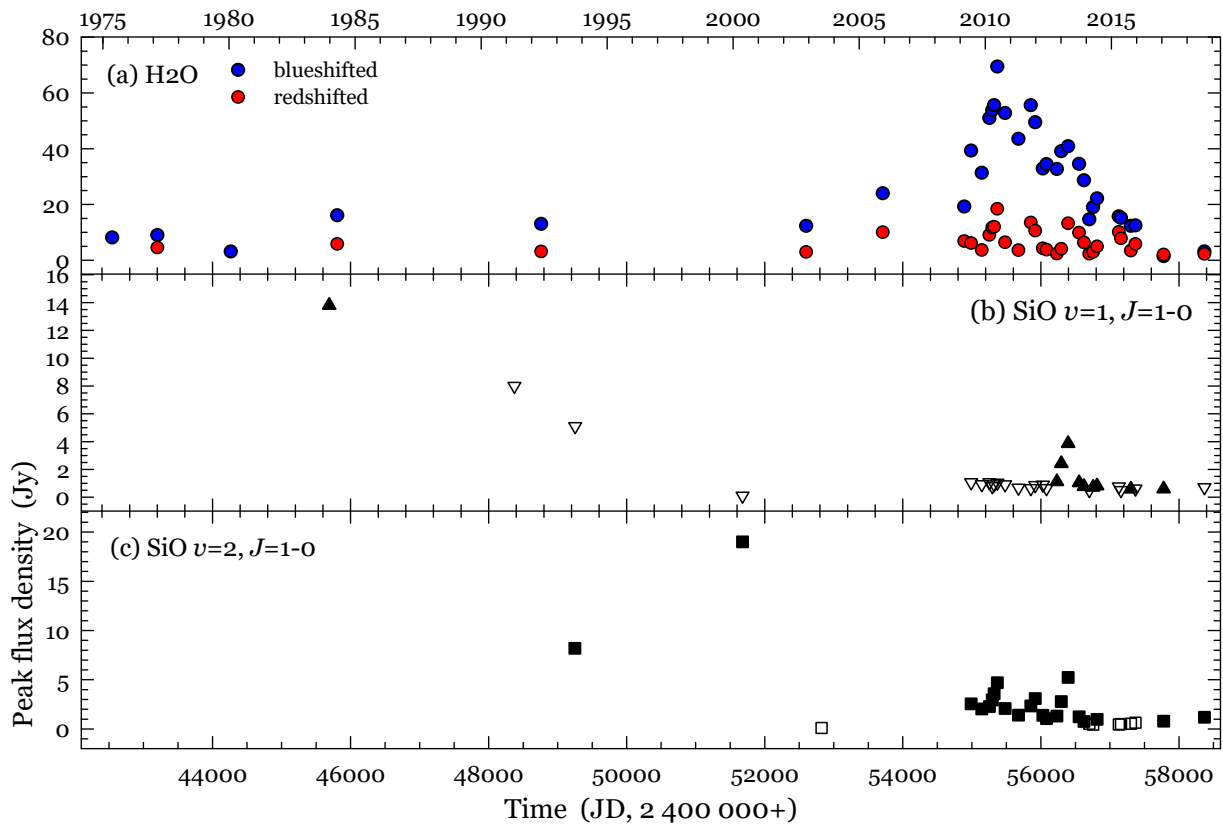}
    \caption{Peak flux densities of the H$_2$O and SiO masers over time in 1975--2017, including the data of other publications (Table~\ref{tab:2}). (a) Flux variations of the blue- and redshifted components of the H$_2$O masers. (b) Same as (a) but for the SiO $v$ = 1, $J$ = 1$\rightarrow$0 maser. A white inverted triangle indicates tentative detection or the upper limit value (3$\sigma$). (c) Same as (a) but for the blueshifted peak flux of the SiO $v$ = 2, $J$ = 1$\rightarrow$0 maser. The peak flux variation of the redshifted components looks a similar trend to that of the blue-shifted one over time. A white square indicates tentative detection or the upper limit (3$\sigma$) value.}
    \label{fig:6}
\end{figure*}

It seems that the H$_2$O flux has increased since about 2000 and peaked at 2010 June (Fig.~\ref{fig:6}a), then, it has decreased steeply. In this case, we could guess that the pulsation-driven shock wave from the central Mira variable has weakened dramatically in recent years, according to the decline of the mass-loss rate associated with the final AGB phase, before forming a PN nucleus. However, the observational results for the 20 or 30 years before our observations are not clearly following this pattern, as shown in Table~\ref{tab:2} and Fig.~\ref{fig:6}a. Although the detected data prior to our monitoring observations are not enough to confirm the variability trend accurately, it seems that the H$_2$O maser intensities have remained silent with an averagely $\sim$\,12\,Jy before clear episode of the intense H$_2$O masing on around 2010. These results leads us to conclude that there has been a transient episode of strong H$_2$O masing on around 2010 (probably denoting a short episode of enhanced mass-loss rate) and a subsequent decline of the mass-loss rate back to values before the burst took place, and this could repeat in the future. Therefore, to establish a valid cause, continuous monitoring observation toward this object is needed.

\subsubsection{SiO masers}

As plotted in Figs.~\ref{fig:4}c and ~\ref{fig:5}b, the SiO $v$ = 1, $J$ = 1$\rightarrow$0 maser was detected in only eight epochs around 2013 April 12. On that date, the SiO $v$ = 1, $J$ = 1$\rightarrow$0 maser shows the strongest intensity, which is comparable to that of the SiO $v$ = 2, $J$ = 1$\rightarrow$0 maser. In addition, as mentioned in Section 3, the SiO $v$ = 1, $J$ = 1$\rightarrow$0 maser appeared only in the blueshifted part, relative to the systemic radial velocity. This means that the SiO $v$ = 1 maser is emitted in a small compact region that shows strong blueshifted velocity components for a limited period. Considering previous records for SiO $v$ = 1 maser detection (Table~\ref{tab:2} and Fig.~\ref{fig:6}b), the equatorial distribution of the SiO maser clumps, and the presence of a secondary star in OH231.8\rq{s} core, the companion might be playing a role in the overall dynamics of the SiO maser region. This brief strong SiO amplification could be related to the passage through the periastron of the binary companion. If this inference is plausible, it suggests that the compact companion has an orbital motion that passes at least close to the turbulent shell between the stellar photosphere and the dust condensation zone ($\sim$\,2--4 $R_{\ast}$ from the primary star) through the periastron, and the SiO $v$ = 1 maser may be associated with short-lived events of enhanced material transfer from the central primary star to the companion star, forming the environment of a high temperature of over 1,000 K and density of over 10$^9$ cm$^{-3}$. \citet{2002AA...385L...1S} estimated $\sim$\,8 mas for the maximum spatial extent of the SiO maser structure around the central Mira variable in the OH231.8 binary core, which, in turn, yields a radius of $\sim9.22\,\times 10^{13}$\,cm at 1.54\,kpc. Thus, the binary separation during the periastron can be estimated to be $\sim$\,6 AU with an elliptical orbit, if it is speculative to attribute the SiO maser flare to the periastron passage of the companion. This maser emission has not been detected or only tentatively detected by several authors since \citeauthor{1991AA...242..211J}\rq{s} detection, other than our result (see Table~\ref{tab:2} and Fig.~\ref{fig:6}b). Therefore, it could be said that the orbit lasts at most $<\,\sim$ 30 yr because other events could have occurred between \citeauthor{1991AA...242..211J}\rq{s} detection and ours. However, this explanation is still speculative. In addition, there is also a high possibility that the SiO maser flare would have nothing to do with the approach of the companion. 

In order to validate the hypothesis for the effect of the companion, it would require a clear detection of a new flare again at any rate. If such a phenomenon occurs again, the effect of the periastron passage of the binary companion may be considered more seriously.

SiO $v$ = 2, $J$ = 1$\rightarrow$0 maser emission was detected relatively frequently in 21 out of 24 epochs, showing 2--4 peaked profile with a strong flux variation by factors of 4--5. The flux variation of each blue- and redshifted peak component and that of the integrated intensity are plotted in Figs.~\ref{fig:4}d and ~\ref{fig:5}b. The SiO $v$ = 2 maser shows a periodic variation for both the peak and integrated flux densities in common with those of the H$_2$O maser. Namely, the flux density variation of the SiO $v$ = 2 maser correlates well with the OLC of the central Mira variable, giving the maximum peak fluxes at phases $\phi$ = 1.15, 2.14, 2.98, i.e. the normalized phase lag of 0--0.15, as for the H$_2$O maser. Moreover, SiO $v$ = 2 maser emission also shows the strongest peak flux density on 2013 April 12, which appears to have been a brief event during our monitoring period, like the SiO $v$ = 1, $J$ = 1$\rightarrow$0 maser emission. The blue- and the redshifted peak components have been consistently detected at similar intensities, and near 2013 April 13, the blueshifted peak component was briefly detected to be two times stronger than the red component. This may be related to the fact that the SiO $v$ = 1 maser also suddenly showed a strong blueshifted peak on the same date. Although the spatial and velocity distributions of SiO $v$ = 1 and 2 masers of known Mira variables are not quite exactly the same, many of them have similar distributions \citep[e.g.][]{2000A&A...360..189D,2005A&A...432..531Y}. Therefore, assuming that the spatial and velocity distributions of SiO $v$ = 1 and 2 masers in OH231.8 are similar, one possible interpretation for sudden strong blueshifted peaks of both masers at the same date is that the companion star could have created a compact environment in which SiO maser emission can be strongly generated (high temperature and density), passing over a circumstellar atmosphere that has a relatively blueshifted velocity with respect to the central star (presumably closest to the periastron), where the velocity gradient may be sufficiently small that maser excitation along radial paths arise.

Meanwhile, the SiO $v$ = 2 maser fluxes show a long-term decrease with time, like those of the H$_2$O maser (Fig.~\ref{fig:5}). As shown in Fig.~\ref{fig:6}c, previously detected fluxes are much stronger than those in our results and more recent fluxes are weaker than our results. As with the case of H$_2$O masers, the decreasing tendency of SiO maser fluxes seems obvious. This decreasing tendency can be also expressed by a simple linear regression analysis as follow (see Fig.~\ref{fig:5}b);
\begin{equation}
f$(MJD)$ = 50.351 - (6.345 \times 10^{-3})\,$MJD$,
\end{equation}
where MJD is the modified Julian date (JD $-$2450000) and $f$(MJD) is the integrated flux density (Jy km s$^{-1}$). The declining slope of the SiO $v$ = 2 maser fluxes is less steep than that of the H$_2$O, however, it is clear that both SiO and H$_2$O maser are losing their strength over time. We conclude that it will not be easy to detect both of these masers from OH231.8 after 2021. This is consistent with the results of the analysis for the H$_2$O maser that the central primary star is currently leaving the AGB phase of stellar evolution \citep{1989ApJ...338..234L,1996A&ARv...7...97H}. This occurs when the gas mass-loss rate decreases drastically, so the circumstellar shell is no longer replenished and becomes detached from the central star.

\subsection{Variations of maser velocity structures}

\begin{figure*}
	\includegraphics[width=170mm]{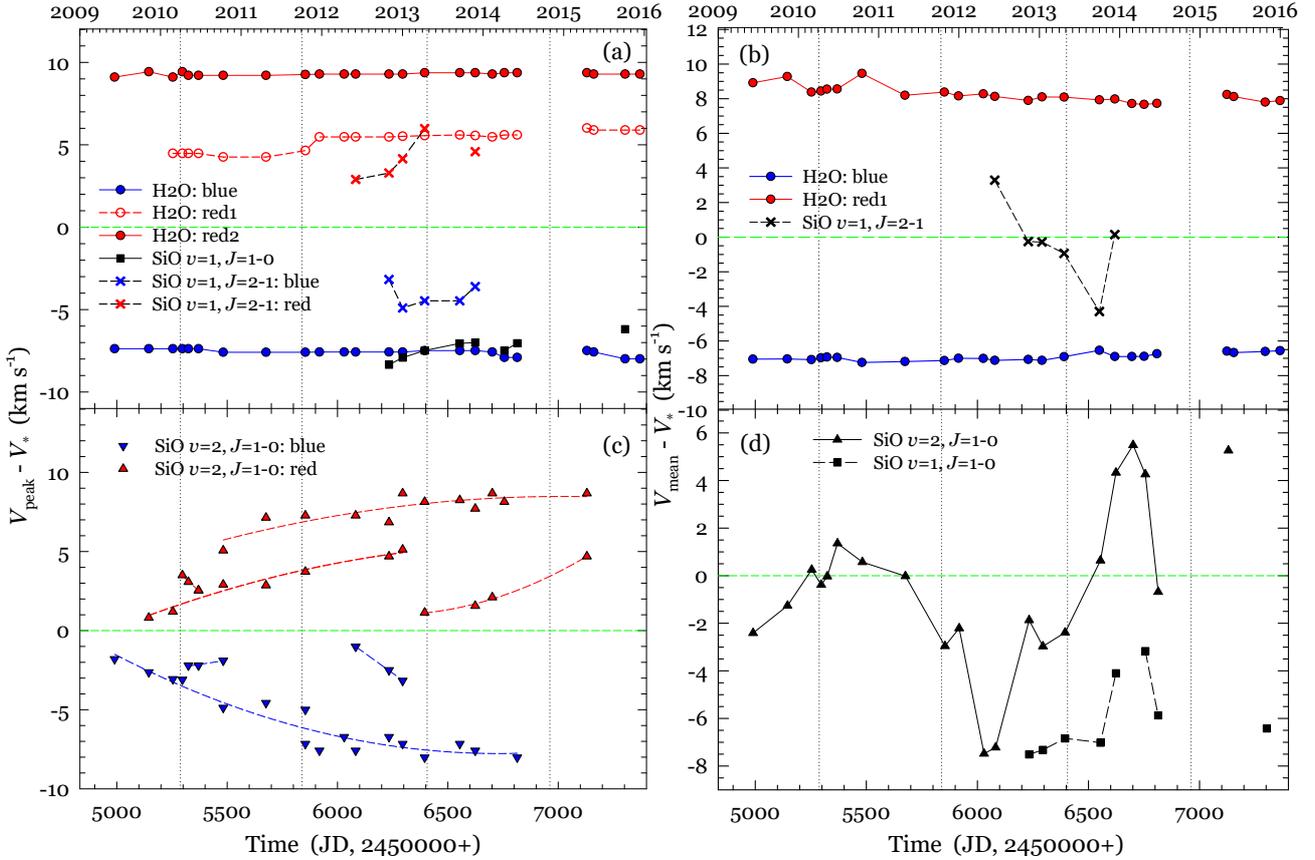}
    \caption{Velocity variation of all detected maser lines as function of time (JD, +2450000). The vertical dotted lines indicate the optical maximum phases and the horizontal green dashed lines correspond to an equal value between the maser and stellar velocity. The symbols used are described in the legend inside the plots. (a): Peak velocity offsets relative to the stellar velocity. (b): Mean velocity offsets of the H$_2$O and SiO $J$ = 2$\rightarrow$1 masers relative to the stellar velocity. Each blue and red circle represents the shifted mean velocity of blue and red components relative to the systemic velocity. (c): Same as (a) but for the dominant 2--4 peaked components presented in Fig.~\ref{fig:b3}. Red and blue dashed lines represent results of a quadratic polynomial regression by the least-squares fitting method. (d): Same as (b) but for the SiO $J$ = 1$\rightarrow$0 masers.}
    \label{fig:7}
\end{figure*}

Many pPNe and PNe have characteristics of a radial outflow whose velocities increase linearly with the distance from the nucleus that one describes as Hubble flows \citep{2002ARA&A..40..439B}. Such kinematic patterns can be a result of ballistic travel, suggesting that all the material we detect was accelerated very fast in the past, but soon reached a terminal velocity. Therefore, we analysed velocity patterns of H$_2$O and SiO masers to tease out long-term dynamic trends of the circumstellar gas in OH231.8. Fig.~\ref{fig:7} shows the peak and mean velocity patterns of detected SiO and H$_2$O maser emission relative to the systematic velocity of the central Mira variable, according to the optical phases. 

\subsubsection{H$_2$O maser}

The peak velocities, which have within the measuring errors ($\approx$\,0.4\,\kms), of the three features of H$_2$O masers relative to the stellar velocity, $V_{\rm peak}-V_{\ast}$, are given in Fig.~\ref{fig:7}a, plotted against observational epoch. Throughout the monitoring program that covers a period over four stellar pulsation cycles, the $V_{\rm peak}$ of the H$_2$O components are remarkably constant within $\pm$0.5\,\kms, distributing in two main groups with the blueshifted components at $-$8.0\,\kms\ and the redshifted components at $\sim$\,5.3 and $\sim$\,9.3\,\kms\ with respect to the $V_{\ast}$. This kinematic pattern of the H$_2$O maser provides an shred of evidence for ballistic travel of bipolar outflow in OH231.8, i.e. that each H$_2$O clump is expanding at a constant velocity, particularly when combined with the similar results from the mean velocity. This can be compared with the measurement of proper motions by \citet{2018MNRAS.476..520D}. No sign of acceleration/deceleration was detected in that study. From these results, we suggest that constant-velocity, ballistic motion, as the most viable paradigm of pPN dynamics, is presented in OH231.8. Quite a number of bipolar pPNe/PNe exhibits such motions, showing cylindrical jets that are accelerated in a way that they mimic a ballistic jet, although the theory is not fully developed \citep[e.g.][]{2001A&A...377..868B,2002ARA&A..40..439B,2004ASPC..313..148C,2007ApJ...663..342H}.

Fig.~\ref{fig:7}b shows mean velocity shifts of each blue- and redshifted part with respect to the stellar velocity, to wit, the bidirectional velocity centroid of the H$_2$O maser, $V_{\rm mean}$(blue)$-V_{\ast}$ and $V_{\rm mean}$(red)$-V_{\ast}$. This graph clearly reflects the results of Fig.~\ref{fig:7}a with variations smaller than $\pm$0.5\,\kms.

\subsubsection{SiO masers}

The peak and mean velocities, $V_{\rm peak}$ and $V_{\rm mean}$, of SiO maser emission with respect to the stellar velocity, $V_{\ast}$, are presented in Fig.~\ref{fig:7} ($V_{\rm peak}-V_{\ast}$ and $V_{\rm mean}-V_{\ast}$). Unfortunately, it is not possible to identify any notable characteristic trend of the SiO $v$ = 1, $J$ = 1$\rightarrow$0 and $J$ = 2$\rightarrow$1 masers, due to the limited detections. On the other hand, those of the SiO $v$ = 2, $J$ = 1$\rightarrow$0 maser emission exhibit a long-term trend, to wit, a gradual increase in the relative peak velocities of both blue- and redshifted components of SiO $v$ = 2, $J$ = 1$\rightarrow$0 maser emission with respect to the $V_{\ast}$ (see Fig.~\ref{fig:7}c). This trend is visible over the OLC cycles of up to 7 years. This long-term velocity pattern is characterized by the polynomial regression using the least squares estimation. Although it is accompanied by a slight decrease and increase in velocity over a period of MJD 5297 -- 5482 (i.e. 2010 April -- October, from $\phi$ = 1.02 to $\phi$ = 1.35, corresponding to a phase of starting stellar contraction), implying a turbulence inside the CSE close to the central Mira variable, a significant and systematic increase of the absolute value of the shifted velocity, $|V_{\rm peak}-V_{\ast}|$, is apparent, giving the asymptotic velocity of $-$7.8\,\kms\ and 8.5\,\kms\ for blue- and redshifted components, respectively. 

From the analysis result, we suggest that the dominant dynamic process in the SiO masing region of OH231.8 is radial acceleration, which is very different from standard AGB stars. The acceleration tends to be higher at the beginning, reaching sometimes an almost constant velocity. During their travels, the SiO masers will quench at some point, as suggested by Fig.~\ref{fig:7}c (all spikes disappear sooner or later).

We note that the asymptotic velocity of SiO $v$ = 2 maser emission is very close to the velocity of H$_2$O peak components, but, does not exceed that of the H$_2$O maser. This indicates that an expansion velocity of the SiO maser clumps may increase closely to the outflow velocity identified in the H$_2$O clumps. However, they will not reach the H$_2$O region spatially, because they expand along the equator, apparently; we guess that no clump will continue masing well before they reach the distance at which the H$_2$O masers are emitting. Thus, SiO maser clumps would be in the short-lived regime of significant acceleration before ballistic dynamics take over.

In Fig.~\ref{fig:7}d, we present the relative mean velocity of SiO $v$ = 2, $J$ = 1$\rightarrow$0 maser emission with respect to the systemic velocity with a black triangle symbol. The relative mean velocity value has changed from being dominantly blueshifted to being dominantly redshifted during the monitoring period. From this pattern, one may guess rotation due to the orbital motion of the central star. However, the most likely cause for the mean velocity pattern of our results could be the changes in the relative peak intensities. The variation of the $V_{\rm mean}$ values seen in Fig.~\ref{fig:7}d cut conspicuous figures in the relatively negative and positive velocities found, respectively, at MJD 6030 and 6701 (2012 April 12 and 2014 February 12). In these two epochs, the maser emission is almost undetected and dominated by only one weak feature, at relatively negative and positive velocities respectively. The variation in the $V_{\rm mean}$ reflects this particular change from emission dominated by the negative velocities to emission with more intense positive spikes, particularly noticeable when the emission is very weak. The velocity of the individual spikes, probably corresponding to the velocity of the maser clumps, does not show any significant changes across these epochs (see Fig.~\ref{fig:7}c). In addition, the $V_{\rm mean}$ of SiO $v$ = 1, $J$ = 2$\rightarrow$1 maser emission show the opposite pattern of change in that of SiO $v$ = 2, $J$ = 1$\rightarrow$0 maser emission, as shown in Fig.~\ref{fig:7}b. Therefore, it is difficult to conclude that this would be caused by rotation. It may be rather related to the unstable conditions and pulsation motions in the SiO maser emitting regions, which will make the same bright SiO spots sometimes being redshifted and sometimes blueshifted. On the other hand, if the clumps are in accelerated expansion, the amplitude of the pattern curve with positive/negative velocity increases over time, reflecting the velocity of the individual spikes, and those with positive/negative velocities would keep with the positive/negative velocity with increasing values of the absolute value. Therefore, we suggest that the most possible mechanism driving the large-scale movement of the SiO maser region is accelerated radial expansion, despite the potential to infer both of the rotation and the expansion of the SiO maser. 

\subsection{Full Width at Zero Power of H$_2$O, SiO masers and SiO thermal line}

Since the SiO $v$ = 2, $J$ = 1$\rightarrow$0 maser of OH231.8 represents the equatorial torus-like structure \citep{2002AA...385L...1S}, it does not appear to have a direct impact on the formation of the high velocity bipolar outflow in this nebula which has been identified from various molecular lines (especially, $^{12}$CO emission). Instead, it seems to be related with the dense equatorial waist of the nebula by the effect of the companion. Measuring high velocity and spatial resolution observations of this SiO maser emission will give us a sense of the expansion and the contraction in the equatorial direction. 

\begin{figure}
    \includegraphics[width=\columnwidth]{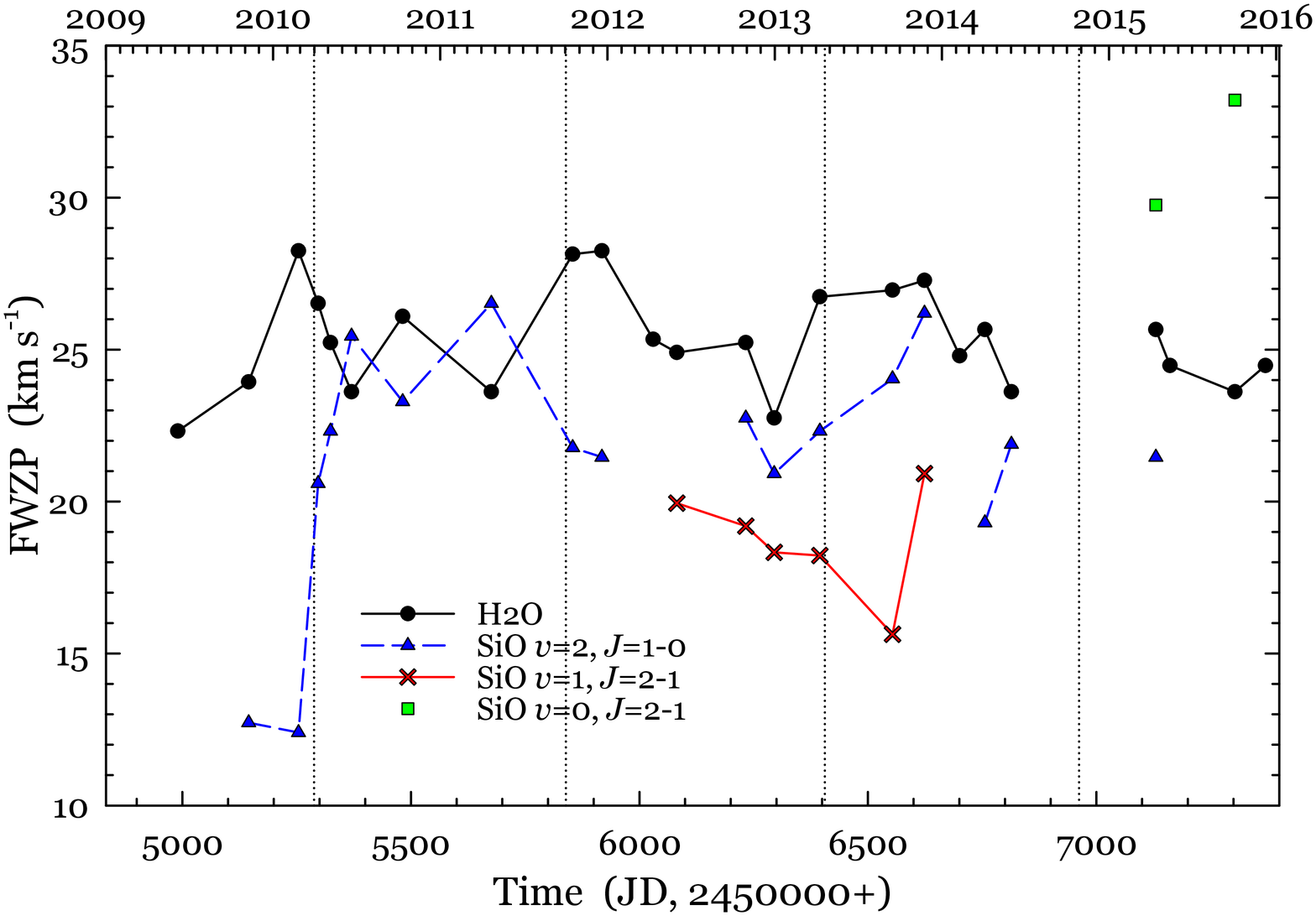}
    \caption{The FWZP variation of H$_2$O and SiO maser lines as function of time (JD, +2450000), with that values of the SiO $v$ = 0, $J$ = 2$\rightarrow$1 thermal line. The symbol used are described in the legend inside the plot.}
    \label{fig:8}
\end{figure}

On the other hand, geometrically, the H$_2$O maser shows the underlying bipolar outflow that probably produce the macro-scale structure of the nebula, which represents the innermost parts of the SiO outflow traed by SiO $J$ = 7$\rightarrow$6 and $^{29}$SiO $J$ = 8$\rightarrow$6 lines \citep{2018A&A...618A.164S}. The velocity extent of the H$_2$O maser line may give the outflow velocity which makes up a sizable fraction of the terminal velocity. Therefore, we measured the full width at zero power (FWZP) of each maser line profile to investigate the outflow/expansion velocities of the H$_2$O and SiO maser regions, together with the SiO $v$ = 0, $J$ = 2$\rightarrow$1 line, and compared with those of other objects in the late stellar evolutionary stage. The FWZP variation of H$_2$O and SiO maser lines over time are presented in Fig.~\ref{fig:8}, with the values of the SiO $v$ = 0, $J$ = 2$\rightarrow$1 thermal line. The average values of the measured FWZPs are given in Table~\ref{tab:3}. Here, those of the SiO $v$ = 1, $J$ = 1$\rightarrow$0 maser were excluded because it only showed the blueshifted peak components. In addition, the FWZPs of the SiO $v$ = 2, $J$ = 1$\rightarrow$0 maser were measured, only where both blue- and redshifted peak components were detected.

As expected from the previous analyses of the peak/mean velocities and Fig.~\ref{fig:8}, the FWZP of the H$_2$O maser line remains relatively stable with a small variation of $\sim$\,6\,\kms, compared to that of SiO maser emission for the entire pulsation cycle during our observation period. The expansion velocity of the H$_2$O maser can be estimated from the velocity difference between the systemic radial velocity of the central star and the most blue- or most redshifted velocity of the H$_2$O maser line. Thus, we estimate the H$_2$O outflow velocity as $\sim$\,18\,\kms. This value is in good agreement with the H$_2$O expansion velocity of $\sim$\,19\,\kms\ estimated by \citet{2018MNRAS.476..520D}. In addition, we also estimate the maximum expansion velocity of $\sim$\,20\,\kms\ of the SiO $v$ = 0, $J$ = 2$\rightarrow$1 thermal emission. There is good agreement between the values obtained from this thermal emission and the H$_2$O maser as a tracer of shocked gas. For reference, the SiO $v$ = 0, $J$ = 7$\rightarrow$6 and $^{29}$SiO $J$ = 8$\rightarrow$7 (thermal) emission lines have been also recently detected to trace a compact bipolar outflow, where the expansion velocities are lower than 35\,\kms\,\citep{2018A&A...618A.164S}. This SiO compact outflow seems to be the large scale counterpart of the H$_2$O maser bipolar flow.

\begin{table}
	\centering
	\caption{The averaged FWZPs ($\Delta v$) of the H$_2$O, SiO masers and SiO thermal line. $\Delta v_{\rm min}$ and $\Delta v_{\rm max}$ representing the minimum and maximum FWZPs, respectively.}
	\label{tab:3}
	\begin{tabular}{cccc}
		\hline
        Lines       & $\Delta v$    & $\Delta v_{\rm min}$ & $\Delta v_{\rm max}$ \\
                    & (km s$^{-1}$) & (km s$^{-1}$)    & (km s$^{-1}$)    \\
		\hline
		H$_2$O                             & 25.3 & 22.3 & 28.3 \\
		SiO $v$ = 2, $J$ = 1$\rightarrow$0 & 21.6 & 12.4 & 26.5 \\
		SiO $v$ = 1, $J$ = 2$\rightarrow$1 & 18.7 & 15.6 & 20.9 \\
		SiO $v$ = 0, $J$ = 2$\rightarrow$1 & 32.8 & 29.8 & 35.8 \\
		\hline
	\end{tabular}
\end{table}

Meanwhile, the FWZP of the SiO $v$ = 2, $J$ = 1$\rightarrow$0 maser shows rapid changes over time, with a large variety of $\sim$\,14\,\kms\ between $\Delta v_{\rm min}$ and $\Delta v_{\rm max}$. This is because the line profile of the SiO $v$ = 2 maser which basically shows double-peaked features, tends to broaden with time, following its peak velocity pattern, suggesting the expansion of the SiO features. Moreover, this emission line traces the torus-like structure in the equatorial direction, so that it cannot directly be compared to the expansion velocity to the estimated values from the H$_2$O maser and SiO thermal line. In other words, according to \citet{2002AA...385L...1S}, \citet{2018MNRAS.476..520D} and our observations, the materials traced by the SiO masers are mainly located and expanded in the equatorial direction, while the outflow materials traced by the H$_2$O maser and SiO thermal line are located and expanded in the axial symmetric bipolar direction.

The equatorial expansion velocity of the SiO $v$ = 2, $J$ = 1$\rightarrow$0 maser also increases gradually over time with the absolute values, $|\rm the~most~\it V_{\rm blue}-V_{\ast}|$ = $\sim$\,7.1--12.0\,\kms\ and $|\rm the~most~\it V_{\rm red}-V_{\ast}|$ = $\sim$\,4.4--16.3\,\kms. The redshifted part shows a more violent velocity change than the blueshifted part, giving the averaged expansion velocity of 11.8\,\kms. The averaged FWZPs presented in Table~\ref{tab:3} and the expansion velocities estimated above are larger than the range of values found in typical Mira variables, and are similar to the values found in OH/IR stars, as presented in \citet{2014AJ....147...22K}. In particular, the averaged FWZP of the H$_2$O maser line, 25.3\,\kms, is very consistent with that of OH/IR stars (25.5\,\kms). In addition, the FWZP ratios of H$_2$O to SiO masers fall within or approach the range of values found in early post-AGB stars \citep{2016JKAS...49..261K}. Therefore, it is presumed that the central host star that powers the maser lines in this nebula is at the end of the AGB phase or is located in the latter half, as also suggested in the previous analyses.

\section{Summary}

We conducted the simultaneous monitoring observations of 22\,GHz H$_2$O maser and  12 other transitions of maser and thermal lines of SiO molecule (43, 86 and 129\,GHz bands) to determine the variability properties of H$_2$O and SiO masers, and to constrain the origin of the peculiar pPN OH231.8, that is powered by a binary system. Both H$_2$O, SiO $v$ = 0, 1, 2, $J$ = 1$\rightarrow$0 and SiO $v$ = 0, 1, $J$ = 2$\rightarrow$1 emissions were simultaneously detected. In addition, we have estimated a new optical period of $\sim$\,554 days, using optical data over the last 10 years (2008--2018) provided by AAVSO. Our monitoring results strongly suggest the periodic variability with an optical phase lag of $\sim$\,0.0--0.15 in especially H$_2$O and SiO $v$ = 2, $J$ = 1$\rightarrow$0 maser lines. In spite of this time delay, the period of the maser flux densities obtained by our monitoring observations agrees with the newly calculated optical period.

The H$_2$O maser emission commonly shows a double-peaked grouping, tracing the bipolar outflow of this object as known from previous works with spatial resolution. The SiO $v$ = 1, $J$ = 1$\rightarrow$0 maser line was detected at only several epochs, with brief strong amplification. We suggest that the latter might be due to a short-lived event of enhanced mass loss, from the central Mira variable, perhaps passing in front of its compact companion, at periastron. Another rotational transition, SiO $v$ = 2, $J$ = 1$\rightarrow$0 maser, shows profiles with 2--4 maser emission peaks at almost every observing epochs. We found a good correlation between the intensity variations and the OLC of the central Mira variable in both H$_2$O and SiO masers, with small phase delay at light maximum. With these characteristics, the intensity of both masers shows a decreasing trend with  time during the monitoring period. It is reasonable to expect that both masers will be finally disappeared in at least 10 years if this tendency is continued. 

We have also found a systematic time-varying behaviour in the velocity structures of the detected masers, which can be used to infer dynamic patterns of the inner circumstellar shells surrounding the central stellar system. The most peaked velocity components of H$_2$O maser profiles appear to be remarkably constant in blue- and redshifted velocity regions without regard to periodic variations, tracing well the bipolar outflow of OH231.8. This is a good evidence of ballistic movement (after a short phase of velocity increase) of the outflow in this nebula, as a paradigm of pPN dynamics, i.e. that the H$_2$O clumps are at present expanding with no significant acceleration, considering spatially expansion of the H$_2$O maser clumps found by \citet{2018MNRAS.476..520D}. On the other hand, the velocity structure of SiO $v$ = 2, $J$ = 1$\rightarrow$0 maser emission shows a general kinematic trend. A systematic increase of the peak velocities of the individual clumps seems to be apparent, showing that its asymptotic velocity is close to a terminal expansion velocity of H$_2$O peak components. The fact that the velocities of the SiO maser clumps tend to converge precisely to the velocity of the H$_2$O maser clumps would indicate that the SiO maser clumps reach that maximum velocity, which would remain constant during a long time in ballistic movement. Namely, if one assumes that the SiO maser clumps are expanding with a velocity that increases with time but with a decreasing acceleration, such that the velocity converges to that of H$_2$O, and that the H$_2$O maser clumps are now in expansion at constant velocity, then, the short phase of acceleration of the ballistic movement of the H$_2$O clumps would be that observed thanks to the SiO-emitting clumps. It is very probable that SiO maser clump motions are more complex than a simple accelerated expansion; however, we suggest that our monitoring results present a trend of expanding dynamics at moderate velocity and show some acceleration. In order to be well documented this phenomenon, also needs a multi-epoch high spatial resolution mapping of the SiO maser. 

By measuring the FWZPs of the H$_2$O, SiO masers and SiO thermal line, we estimated the expansion velocity of bipolar outflow (traced by the H$_2$O maser and the SiO thermal emission) and the equatorial expansion velocity (traced by the SiO $v$ = 2, $J$ = 1$\rightarrow$0 maser). 

From all of our analyses, we speculate that the central host star of OH231.8 is close to the tip of the AGB phase, and that the mass-loss rate recently started to decrease because of incipient post-AGB evolution. In addition, if H$_2$O maser flux history in the 1975--2018 period, is interpreted as a reflection of the mass-loss rate, then the strong brief episode of H$_2$O masing in 2010 could imply a last period of intense mass-loss before a gradual decline at the end of the AGB phase.

\section*{Acknowledgements}

We are grateful to the anonymous referee of this paper for his/her helpful comments. The KVN observations are supported through the high-speed network connections among the KVN sites provided by the KREONET (Korea Research Environment Open Network), which is managed and operated by the KISTI (Korea Institute of Science and Technology Information). We acknowledge with thanks the variable star observations from the AAVSO International Database contributed by observers worldwide and used in this research. This work has been supported by the Spanish MINECO, grant AYA2016-78994-P, and partly sponsored by the 100 Talents Project of the Chinese Academy of Sciences, the National Natural Science Foundation of China under grant 11673051.



\bibliographystyle{mnras}
\bibliography{Reference}



\appendix
\section{Tables}
For each molecular transition, the maximum antenna temperature, the rms noise, the integrated antenna temperature, the velocities at the flux maximum and the intensity-weighted mean are given in Columns 2 to 6. The observed dates and their corresponding phases of the OLC (maximum light = 0.0, from $\phi$ = 0.46 to 4.74) for QX~Pup are listed in Column 7. The estimate of the optical magnitudes is provided by the American Association of Variable Star Observers (AAVSO). The conversion factors and telescopes used for observations are listed in Columns 8 and 9. For the SiO $v$ = 0, $J$ = 2$\rightarrow$1 thermal emission of Table~\ref{tab:a2}, we fitted a Gaussian function, giving in parentheses. Table~\ref{tab:a3} shows the negative results.

\begin{table*}
	\centering
	\caption{Parameters of the H$_2$O and SiO maser lines detected toward OH231.8.}
	\label{tab:a1}
	\begin{tabular}{ccccccccc}
		\hline
Molecular & $T_{\rm A}^{\ast}$(peak) & rms & \x $\int\! T_{\rm A}^{\ast}dv$\e & $V_{\rm peak}$ & $V_{\rm mean}$ & Date(phase) & C.F.$^a$ & Obs.$^b$ \\
Transition & (K) & (K) & (K\,km\,s$^{-1}$) & (km\,s$^{-1}$) & (km\,s$^{-1}$) & (yymmdd) & (Jy\,K$^{-1}$) &     \\
(1)        & (2) & (3) & (4)               & (5)            & (6)            & (7)      & (8)            & (9) \\
		\hline
H$_2$O 6$_{1,6}$--5$_{2,3}$ 
& 3.20, $\cdots$, 0.50 & 0.05 & 5.63  & 25.6, $\cdots$, 42.1 & 27.7 & 090607(0.46) & 12.26 & YS \\
& 2.56, $\cdots$, 0.30 & 0.03 & 4.01  & 25.6, $\cdots$, 42.4 & 28.0 & 091109(0.74) & 12.26 & YS \\
& 4.16, 0.33,     0.74 & 0.03 & 7.97  & 25.6, 37.5,     42.1 & 28.5 & 100226(0.94) & 12.26 & YS \\
& 4.39, 0.28,     0.95 & 0.02 & 9.08  & 25.6, 37.5,     42.4 & 28.9 & 100410(1.02) & 12.26 & YS \\
& 4.60, 0.29,     0.99 & 0.02 & 9.60  & 25.6, 37.5,     42.2 & 29.0 & 100507(1.07) & 12.08 & TN \\
& 5.66, 0.39,     1.50 & 0.04 & 12.13 & 25.6, 37.5,     42.2 & 29.1 & 100622(1.15) & 12.26 & YS \\
& 4.37, 0.10,     0.53 & 0.04 & 8.21  & 25.4, 37.3,     42.2 & 27.4 & 101012(1.35) & 12.08 & TN \\
& 3.55, 0.09,     0.29 & 0.02 & 5.96  & 25.4, 37.3,     42.2 & 27.0 & 110424(1.70) & 12.26 & YS \\
& 4.53, 0.13,     1.10 & 0.02 & 9.81  & 25.4, 37.7,     42.3 & 28.4 & 111019(2.03) & 12.26 & YS \\
& 4.03, 0.19,     0.86 & 0.01 & 8.50  & 25.4, 38.5,     42.3 & 28.8 & 111222(2.14) & 12.26 & YS \\
& 2.68, 0.15,     0.35 & 0.02 & 5.55  & 25.4, 38.5,     42.3 & 27.5 & 120412(2.34) & 12.26 & YS \\
& 2.81, 0.10,     0.30 & 0.02 & 5.39  & 25.4, 38.5,     42.3 & 27.2 & 120603(2.43) & 12.26 & YS \\
& 2.26, 0.13,     0.16 & 0.01 & 4.62  & 25.4, 38.5,     42.3 & 27.7 & 121101(2.69) & 14.49 & YS \\
& 2.94, 0.22,     0.30 & 0.01 & 5.58  & 25.4, 38.5,     42.3 & 27.5 & 130102(2.80) & 13.29 & US \\
& 3.07, 0.48,     0.99 & 0.02 & 7.88  & 25.5, 38.6,     42.4 & 29.5 & 130412(2.98) & 13.29 & US \\
& 2.42, 0.56,     0.69 & 0.03 & 6.90  & 25.5, 38.6,     42.4 & 30.3 & 130918(3.27) & 14.24 & YS \\ 
& 2.01, 0.40,     0.45 & 0.01 & 4.83  & 25.5, 38.6,     42.4 & 29.5 & 131127(3.39) & 14.24 & YS \\
& 1.12, 0.20,     0.18 & 0.02 & 2.63  & 25.4, 38.5,     42.3 & 28.6 & 140212(3.53) & 13.07 & US \\
& 1.34, 0.23,     0.21 & 0.01 & 3.08  & 25.1, 38.6,     42.4 & 28.4 & 140408(3.63) & 14.24 & YS \\
& 1.61, 0.36,     0.36 & 0.03 & 4.03  & 25.1, 38.6,     42.4 & 28.8 & 140605(3.73) & 13.75 & TN \\
& 1.08, 0.34,     0.70 & 0.01 & 3.73  & 25.5, 39.0,     42.4 & 31.5 & 150417(4.31) & 14.49 & YS \\
& 1.05, 0.40,     0.54 & 0.02 & 3.28  & 25.4, 38.9,     42.3 & 31.1 & 150518(4.36) & 14.49 & YS \\
& 0.92, 0.25,     0.26 & 0.01 & 2.72  & 25.0, 38.9,     42.3 & 29.1 & 151007(4.62) & 13.29 & TN \\
& 0.86, 0.33,     0.40 & 0.02 & 2.99  & 25.0, 38.9,     42.3 & 30.6 & 151213(4.74) & 14.49 & YS \\
[1.2mm]
$^{28}$SiO $v$ = 1, $J$ = 1$\rightarrow$0 
& $\cdots$ & 0.03 & $\cdots$ & $\cdots$ & $\cdots$ & 090607(0.46) & 11.90 & YS \\
& $\cdots$ & 0.03 & $\cdots$ & $\cdots$ & $\cdots$ & 091109(0.74) & 11.90 & YS \\
& $\cdots$ & 0.03 & $\cdots$ & $\cdots$ & $\cdots$ & 100226(0.94) & 11.90 & YS \\
& $\cdots$ & 0.02 & $\cdots$ & $\cdots$ & $\cdots$ & 100410(1.02) & 11.90 & YS \\
& $\cdots$ & 0.02 & $\cdots$ & $\cdots$ & $\cdots$ & 100507(1.07) & 13.29 & TN \\
& $\cdots$ & 0.03 & $\cdots$ & $\cdots$ & $\cdots$ & 100622(1.15) & 11.90 & YS \\
& $\cdots$ & 0.02 & $\cdots$ & $\cdots$ & $\cdots$ & 101012(1.35) & 13.29 & TN \\
& $\cdots$ & 0.02 & $\cdots$ & $\cdots$ & $\cdots$ & 110424(1.70) & 11.90 & YS \\
& $\cdots$ & 0.02 & $\cdots$ & $\cdots$ & $\cdots$ & 111019(2.03) & 11.90 & YS \\
& $\cdots$ & 0.02 & $\cdots$ & $\cdots$ & $\cdots$ & 111222(2.14) & 11.90 & YS \\
& $\cdots$ & 0.03 & $\cdots$ & $\cdots$ & $\cdots$ & 120412(2.34) & 11.90 & YS \\
& $\cdots$ & 0.02 & $\cdots$ & $\cdots$ & $\cdots$ & 120603(2.43) & 11.90 & YS \\
& 0.08     & 0.01 & 0.20     & 24.7     & 25.5     & 121101(2.69) & 13.29 & YS \\
& 0.17     & 0.01 & 0.31     & 25.1     & 25.7     & 130102(2.80) & 14.24 & US \\
& 0.27     & 0.01 & 0.54     & 25.5     & 26.2     & 130412(2.98) & 14.24 & US \\
& 0.08     & 0.01 & 0.18     & 26.0     & 26.0     & 130918(3.27) & 12.46 & YS \\
& 0.05     & 0.01 & 0.22     & 26.0     & 28.9     & 131127(3.39) & 12.46 & YS \\
& $\cdots$ & 0.01 & $\cdots$ & $\cdots$ & $\cdots$ & 140212(3.53) & 12.86 & US \\
& 0.06     & 0.01 & 0.26     & 25.5     & 29.8     & 140408(3.63) & 12.46 & YS \\
& 0.06     & 0.02 & 0.23     & 26.0     & 25.9     & 140605(3.73) & 13.29 & TN \\
& $\cdots$ & 0.01 & $\cdots$ & $\cdots$ & $\cdots$ & 150417(4.31) & 12.65 & YS \\
& 0.05     & 0.01 & 0.06     & 26.8     & 26.6     & 151007(4.62) & 12.65 & TN \\
& $\cdots$ & 0.02 & $\cdots$ & $\cdots$ & $\cdots$ & 151213(4.74) & 12.65 & YS \\
[1.2mm]
$^{28}$SiO $v$ = 2, $J$ = 1$\rightarrow$0 
& 0.22, $\cdots$ & 0.04 & 0.53 & 31.2, $\cdots$ & 30.6 & 090607(0.46) & 11.90 & YS \\
& 0.17, 0.20     & 0.03 & 0.63 & 30.4, 33.8     & 31.7 & 091109(0.74) & 11.90 & YS \\
& 0.19, 0.24     & 0.03 & 1.05 & 29.9, 34.2     & 33.3 & 100226(0.94) & 11.90 & YS \\
& 0.25, 0.23     & 0.02 & 1.81 & 29.9, 36.5     & 32.6 & 100410(1.02) & 11.90 & YS \\
& 0.27, 0.23     & 0.02 & 1.90 & 30.8, 36.1     & 33.0 & 100507(1.07) & 13.29 & TN \\
& 0.39, 0.41     & 0.03 & 2.74 & 30.8, 35.5     & 34.4 & 100622(1.15) & 11.90 & YS \\
& 0.16, 0.34     & 0.02 & 1.29 & 28.0, 35.9     & 33.6 & 101012(1.35) & 13.29 & TN \\
& 0.12, 0.12     & 0.02 & 0.90 & 28.4, 35.9     & 33.0 & 110424(1.70) & 11.90 & YS \\
& 0.20, 0.14     & 0.02 & 1.25 & 25.8, 36.7     & 30.0 & 111019(2.03) & 11.90 & YS \\
& 0.26, 0.21     & 0.02 & 1.64 & 25.4, 40.3     & 30.8 & 111222(2.14) & 11.90 & YS \\
& 0.12, $\cdots$ & 0.03 & 0.37 & 26.3, $\cdots$ & 25.5 & 120412(2.34) & 11.90 & YS \\
& 0.09, $\cdots$ & 0.02 & 0.14 & 25.4, $\cdots$ & 25.8 & 120603(2.43) & 11.90 & YS \\
& 0.10, 0.06     & 0.01 & 0.60 & 26.3, 40.3     & 31.1 & 121101(2.69) & 13.29 & YS \\
& 0.20, 0.11     & 0.01 & 0.99 & 25.8, 37.7     & 30.0 & 130102(2.80) & 14.24 & US \\
		\hline
	\end{tabular}
\end{table*}

\begin{table*}
	\centering
	\contcaption{.}
	\label{tab:continued}
	\begin{tabular}{ccccccccc}
		\hline
Molecular & $T_{\rm A}^{\ast}$(peak) & rms & \x $\int\! T_{\rm A}^{\ast}dv$\e & $V_{\rm peak}$ & $V_{\rm mean}$ & Date(phase) & C.F.$^a$ & Obs.$^b$ \\
Transition & (K) & (K) & (K\,km\,s$^{-1}$) & (km\,s$^{-1}$) & (km\,s$^{-1}$) & (yymmdd) & (Jy\,K$^{-1}$) &     \\
(1)        & (2) & (3) & (4)               & (5)            & (6)            & (7)      & (8)            & (9) \\
		\hline			   						   
& 0.37,     0.20     & 0.01 & 1.70     & 29.8,     38.1     & 30.6     & 130412(2.98) & 14.24 & US \\
& 0.10,     0.15     & 0.01 & 0.99     & 25.8,     41.2     & 33.6     & 130918(3.27) & 12.46 & YS \\
& 0.06,     0.10     & 0.01 & 0.88     & 25.4,     41.2     & 37.3     & 131127(3.39) & 12.46 & YS \\
& $\cdots$, 0.08     & 0.01 & 0.26     & $\cdots$, 40.7     & 38.5     & 140212(3.53) & 12.86 & US \\
& $\cdots$, 0.10     & 0.01 & 0.51     & $\cdots$, 35.1     & 37.3     & 140408(3.63) & 12.46 & YS \\
& 0.07,     0.07     & 0.02 & 0.53     & 25.0,     41.1     & 32.3     & 140605(3.73) & 13.29 & TN \\
& $\cdots$, 0.11     & 0.01 & 0.64     & $\cdots$, 37.7     & 38.3     & 150417(4.31) & 12.65 & YS \\
& $\cdots$           & 0.01 & $\cdots$ & $\cdots$           & $\cdots$ & 151007(4.62) & 12.65 & TN \\
& $\cdots$           & 0.02 & $\cdots$ & $\cdots$           & $\cdots$ & 151213(4.74) & 12.65 & YS \\
[1.2mm]
$^{28}$SiO $v$ = 1, $J$ = 2$\rightarrow$1 
& $\cdots$, 0.09 & 0.02 & 0.31     & $\cdots$, 35.9 & 36.3      & 120603(2.43) & 13.99 & YS \\
& 0.06,     0.07 & 0.01 & 0.41     & 29.8,     36.3 & 32.8      & 121101(2.69) & 13.99 & YS \\
& 0.05,     0.05 & 0.01 & 0.31     & 28.1,     37.2 & 32.7      & 130102(2.80) & 17.33 & US \\
& 0.10,     0.06 & 0.02 & 0.60     & 28.5,     39.0 & 32.1      & 130412(2.98) & 17.33 & US \\
& 0.10, $\cdots$ & 0.02 & 0.35     & 28.5, $\cdots$ & 28.7      & 130918(3.27) & 16.27 & YS \\
& 0.07,     0.07 & 0.01 & 0.38     & 29.4,     37.6 & 33.2      & 131127(3.39) & 16.27 & YS \\
& $\cdots$       & 0.02 & $\cdots$ &	$\cdots$    & $\cdots$	& 140212(3.53) & 17.72 & US \\
& $\cdots$       & 0.01 & $\cdots$ &	$\cdots$    & $\cdots$	& 140408(3.63) & 16.27 & YS \\
& $\cdots$       & 0.03 & $\cdots$ &	$\cdots$    & $\cdots$	& 140605(3.73) & 16.61 & TN \\
& $\cdots$       & 0.01 & $\cdots$ &	$\cdots$    & $\cdots$	& 150417(4.31) & 15.04 & YS \\
& $\cdots$       & 0.01 & $\cdots$ &	$\cdots$    & $\cdots$	& 151007(4.62) & 15.94 & TN \\
& $\cdots$       & 0.02 & $\cdots$ &	$\cdots$    & $\cdots$	& 151213(4.74) & 15.04 & YS \\
 		\hline
\multicolumn{9}{l}{$^a$ indicates the conversion factor between the antenna temperature and the flux density.}\\
\multicolumn{9}{l}{$^b$ indicates telescopes used for observation (YS: KVN Yonsei telescope, US: KVN Ulsan telescope, TN: KVN Tamna telescope).}\\
	\end{tabular}
\end{table*}

\begin{table*}
	\centering
	\caption{Parameters of SiO $v$ = 0 emission detected toward OH231.8. The values in parentheses show the results fitted with a Gaussian function.}
	\label{tab:a2}
	\begin{tabular}{ccccccccc}
		\hline
Molecular & $T_{\rm A}^{\ast}$(peak) & rms & \x $\int\! T_{\rm A}^{\ast}dv$\e & $V_{\rm peak}$ & $V_{\rm mean}$ & Date(phase) & C.F. & Obs. \\
Transition & (K) & (K) & (K\,km\,s$^{-1}$) & (km\,s$^{-1}$) & (km\,s$^{-1}$) & (yymmdd) & (Jy\,K$^{-1}$) &     \\
(1)        & (2) & (3) & (4)               & (5)            & (6)            & (7)      & (8)            & (9) \\
		\hline
$^{28}$SiO $v$ = 0, $J$ = 1$\rightarrow$0 
                               & 0.09 & 0.02 & 0.32 & 25.4 & 25.2 & 120603(2.43) & 11.90 & YS \\
					           & 0.12 & 0.02 & 0.65 & 25.8 & 25.5 & 150518(4.36) & 12.65 & YS \\
						       & 0.05 & 0.02 & 0.10 & 25.8 & 25.3 & 151007(4.62) & 12.65 & TN \\
						       & 0.08 & 0.02 & 0.25 & 26.3 & 26.1 & 151213(4.74) & 12.65 & YS \\
						       [1.2mm] 
$^{28}$SiO $v$ = 0, $J$ = 2$\rightarrow$1 
& 0.12     & 0.02   & 1.16     & 31.6     & 31.5     & 150501(4.33) & 16.61 & US \\
& (0.06)   &        & (1.19)   & (33.1)   &          &              &       &    \\		
& 0.09     & 0.01   & 1.31     & 30.7     & 32.6     & 151007(4.62) & 15.94 & TN \\
& (0.06)   &        & (1.35)   & (33.2)   &          &              &       &    \\
& $\cdots$ & 0.02   & $\cdots$ & $\cdots$ & $\cdots$ & 151213(4.74) & 15.04 & YS \\
 		\hline
	\end{tabular}
\end{table*}

\begin{table}
	\centering
	\caption{Negative results}
	\label{tab:a3}
	\begin{tabular}{cccc}
		\hline
		Molecular  & rms & Date(phase) & Obs.\\
		Transition & (K) & (yymmdd)    &     \\
		\hline
$^{28}$SiO $v$ = 3, $J$ = 1$\rightarrow$0 & 0.01 & 150501(4.33) & US \\
                                          & 0.01 & 151008(4.62) & US \\
                                          & 0.01 & 151215(4.75) & YS \\
                               [1.2mm]
$^{28}$SiO $v$ = 4, $J$ = 1$\rightarrow$0 & 0.01 & 150501(4.33) & US \\
                                          & 0.01 & 151008(4.62) & US \\
                                          & 0.01 & 151215(4.75) & YS \\
                               [1.2mm]
$^{28}$SiO $v$ = 2, $J$ = 2$\rightarrow$1 & 0.02 & 120603(2.43) & YS \\
                                          & 0.02 & 150518(4.36) & YS \\
                                          & 0.02 & 151008(4.62) & US \\
                                          & 0.02 & 151215(4.75) & YS \\
                               [1.2mm]
$^{28}$SiO $v$ = 0, $J$ = 3$\rightarrow$2 & 0.03 & 150501(4.33) & US \\
                                          & 0.02 & 151007(4.62) & TN \\
                                          & 0.03 & 151213(4.74) & YS \\
$^{28}$SiO $v$ = 1, $J$ = 3$\rightarrow$2 & 0.03 & 120603(2.43) & YS \\
                                          & 0.02 & 130102(2.80) & US \\
                                          & 0.02 & 130412(2.98) & US \\
                                          & 0.04 & 130918(3.27) & YS \\
                                          & 0.02 & 131127(3.39) & YS \\
                                          & 0.02 & 140408(3.63) & YS \\
                                          & 0.03 & 140605(3.73) & TN \\
                                          & 0.01 & 150417(4.31) & YS \\
                                          & 0.02 & 151007(4.62) & TN \\
                                          & 0.02 & 151213(4.74) & YS \\
                               [1.2mm]
$^{28}$SiO $v$ = 2, $J$ = 3$\rightarrow$2 & 0.03 & 120603(2.43) & YS \\
                                          & 0.04 & 150518(4.36) & YS \\
                                          & 0.03 & 151008(4.62) & US \\
                                          & 0.02 & 151215(4.75) & YS \\
                               [1.2mm]
$^{29}$SiO $v$ = 0, $J$ = 1$\rightarrow$0 & 0.03 & 100226(0.94) & YS \\
                                          & 0.02 & 100410(1.02) & YS \\
                                          & 0.02 & 100507(1.07) & TN \\
                                          & 0.03 & 100622(1.15) & YS \\
                                          & 0.03 & 101012(1.35) & TN \\
                                          & 0.02 & 110424(1.70) & YS \\
                                          & 0.02 & 111019(2.03) & YS \\
                                          & 0.02 & 111222(2.14) & YS \\
                                          & 0.02 & 120412(2.34) & YS \\
		\hline
	\end{tabular}
\end{table}

\section{Figures}
\begin{figure}
   \includegraphics[width=\columnwidth]{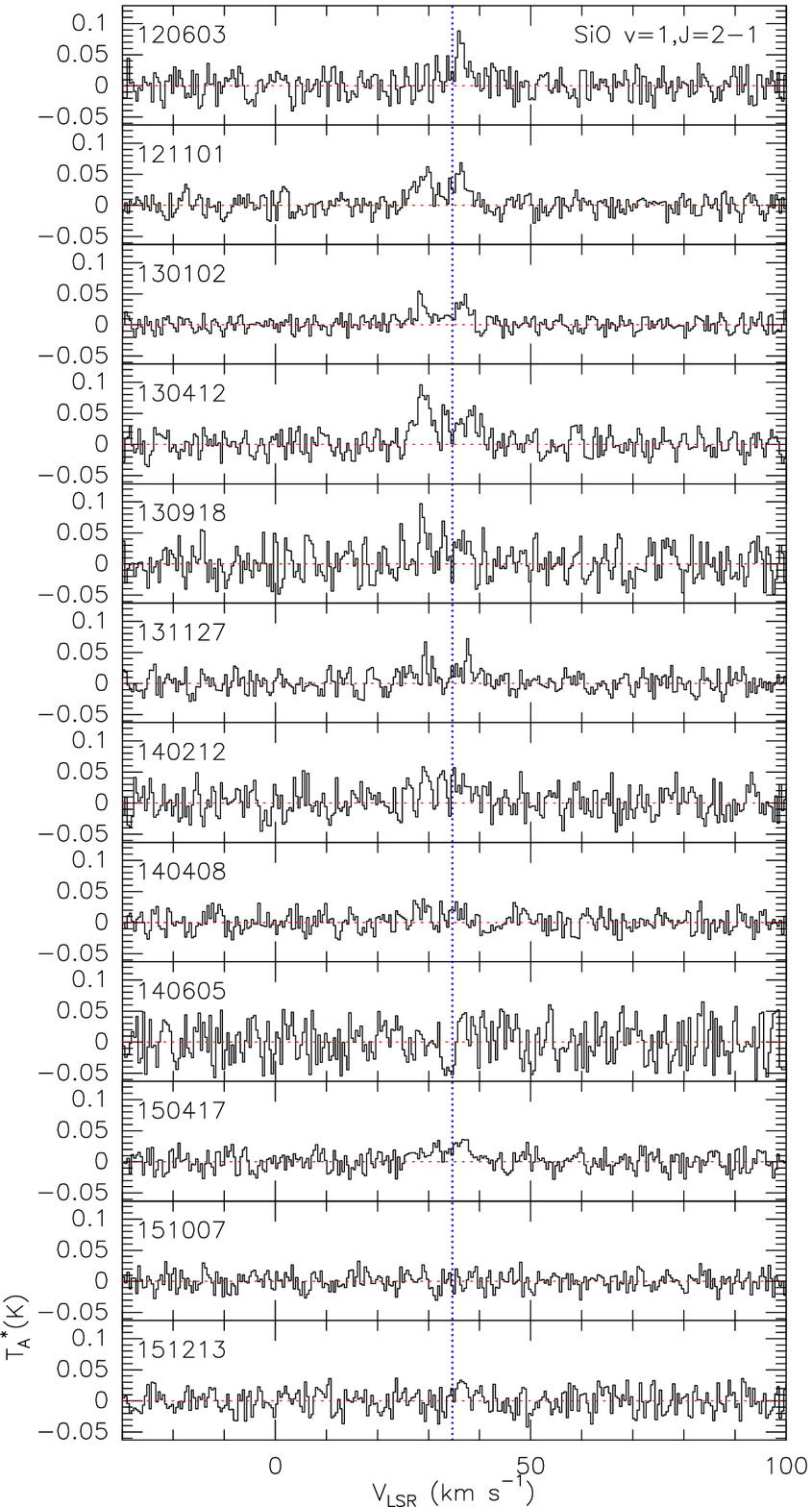}
   \vspace{-5mm}
   \caption{spectra of the SiO $v$ = 1, $J$ = 2$\rightarrow$1 maser monitored toward OH231.8.}
   \label{fig:b4}
\end{figure}

\begin{figure}
   \includegraphics[width=\columnwidth]{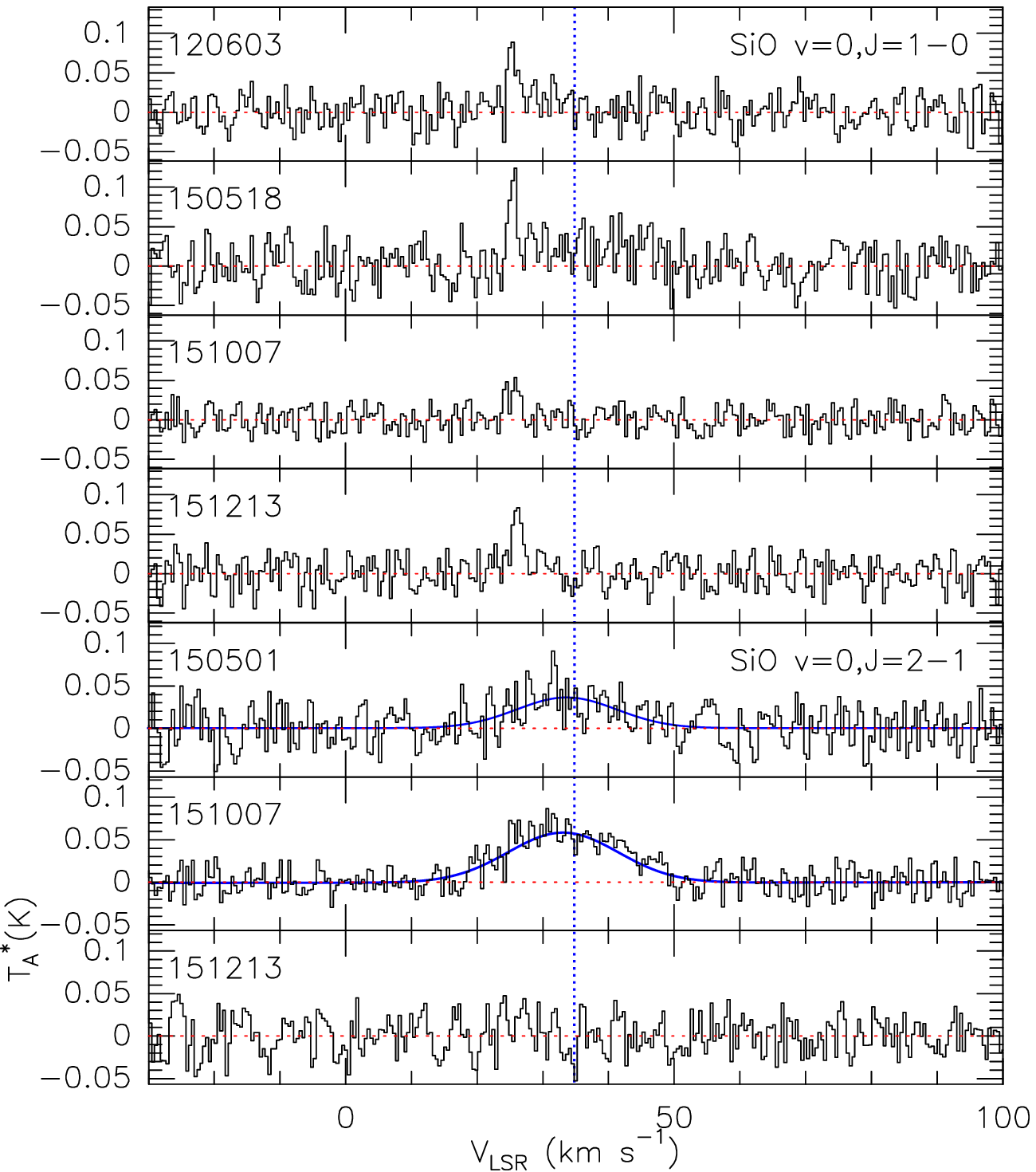}
   \vspace{-5mm}
   \caption{spectra of the SiO $v$ = 0, $J$ = 1$\rightarrow$0 and 2$\rightarrow$1 lines monitored toward OH231.8.}
   \label{fig:b5}
\end{figure}

Figs.~\ref{fig:b1}--~\ref{fig:b5} show the line profiles of all detected transitions with the observed date. The dates and maser transitions are indicated in each spectrum. The intensity is given in units of the antenna temperature $T_{\rm A}^{\ast}$ (K), and the abscissa is $V_{\rm LSR}$ (\kms). The vertical blue dotted lines mark the systemic radial velocity ($\sim$\,35\,\kms) with respect to the local standard of rest (LSR).

\bsp	
\label{lastpage}
\clearpage

\begin{figure}
   \includegraphics[width=\columnwidth]{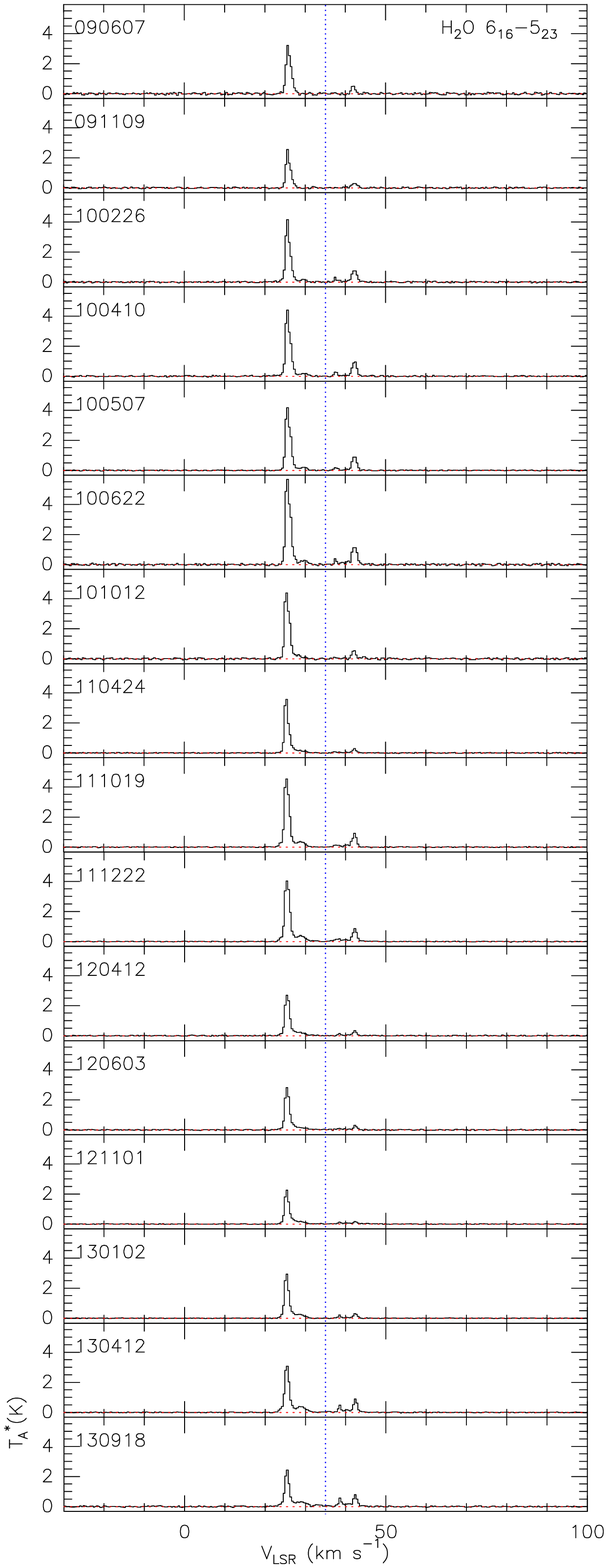}
   \vspace{-5mm}
   \caption{spectra of the H$_2$O maser monitored toward OH231.8.}
   \label{fig:b1}
\end{figure}

\begin{figure}
   \includegraphics[width=\columnwidth]{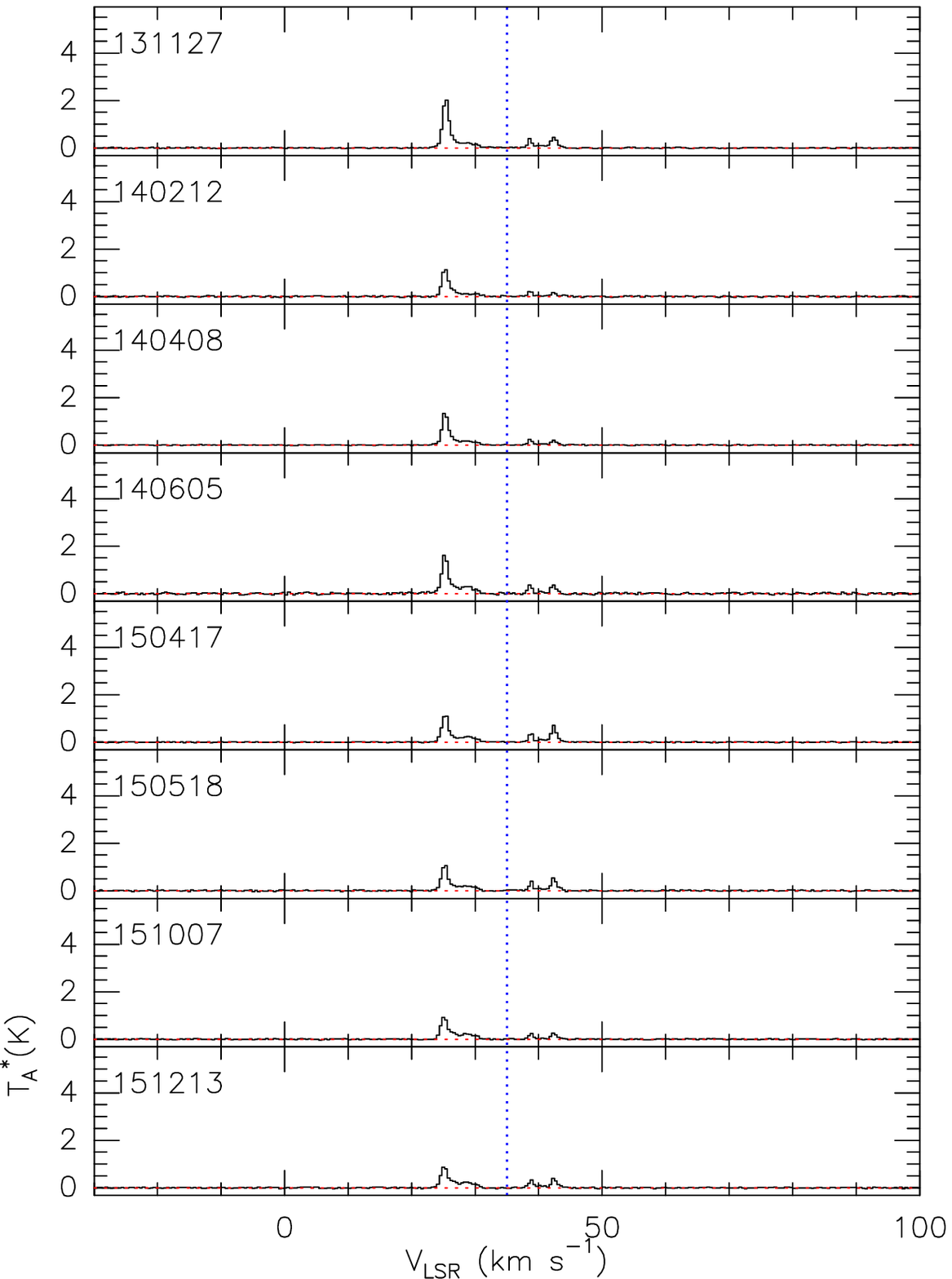}
   \vspace{-5mm}
   \contcaption{.}
   \label{fig:continued}
\end{figure}

\begin{figure}
   \includegraphics[width=\columnwidth]{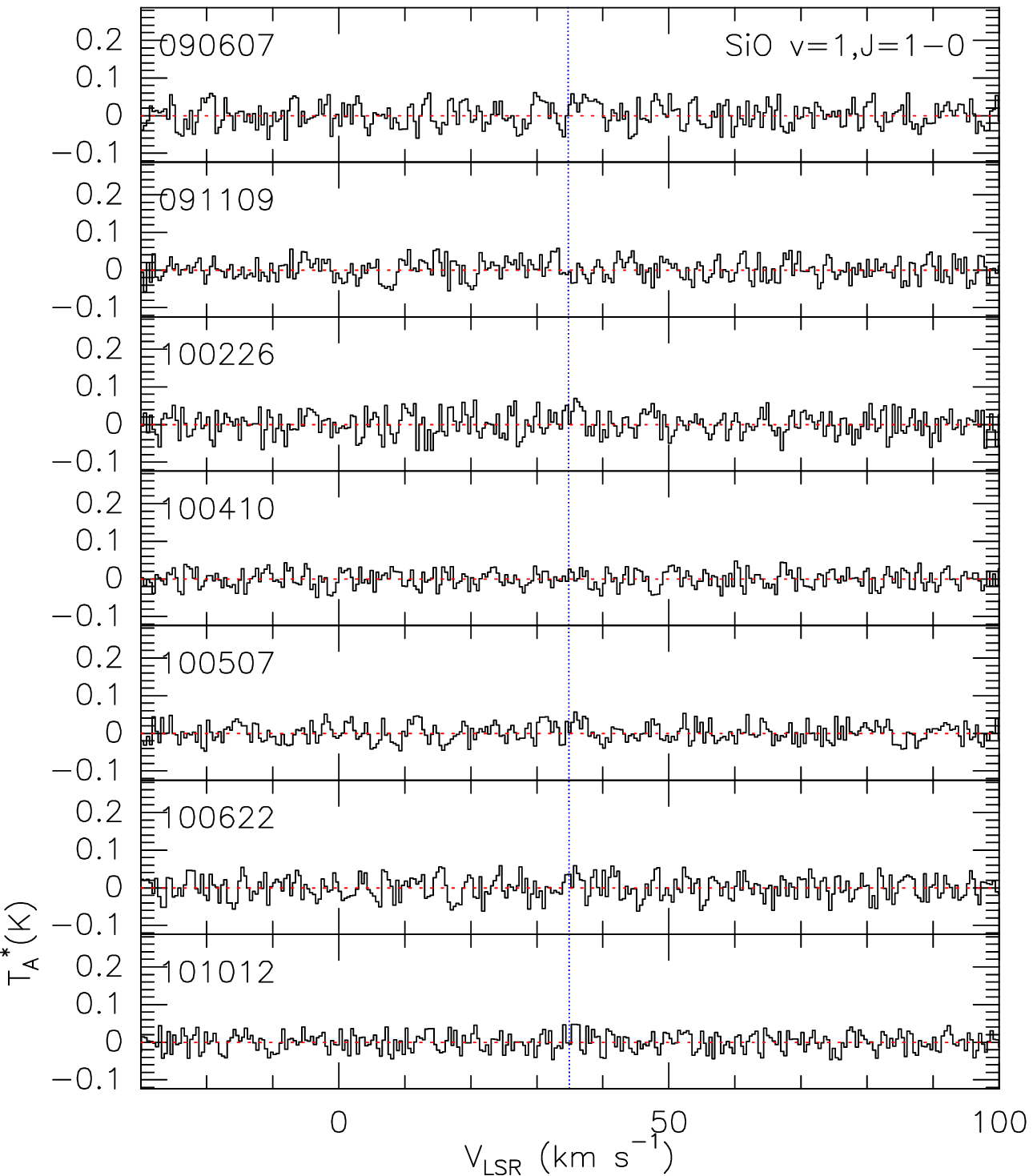}
   \vspace{-5mm}
   \caption{spectra of the SiO $v$ = 1, $J$ = 1$\rightarrow$0 maser monitored toward OH231.8.}
   \label{fig:b2}
\end{figure}

\begin{figure}
   \includegraphics[width=\columnwidth]{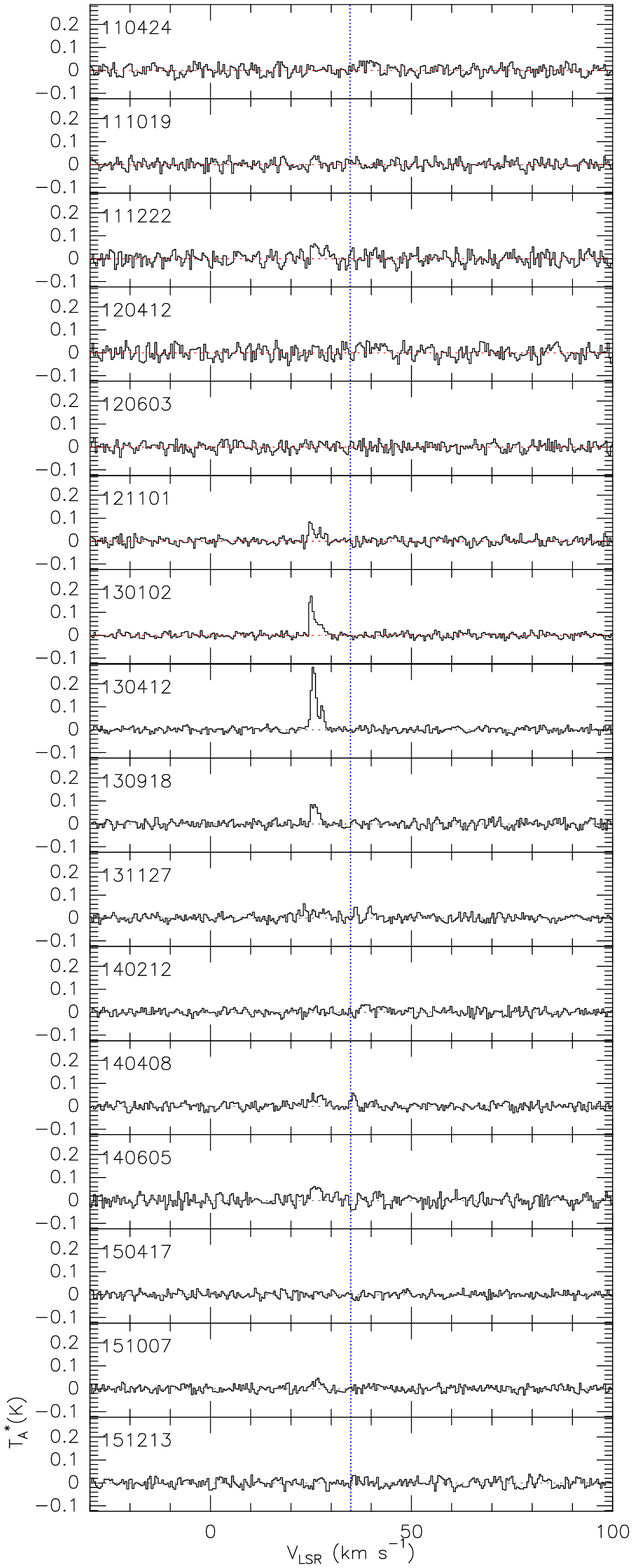}
   \vspace{-5mm}
   \contcaption{.}
   \label{fig:continued}
\end{figure}

\begin{figure}
   \includegraphics[width=\columnwidth]{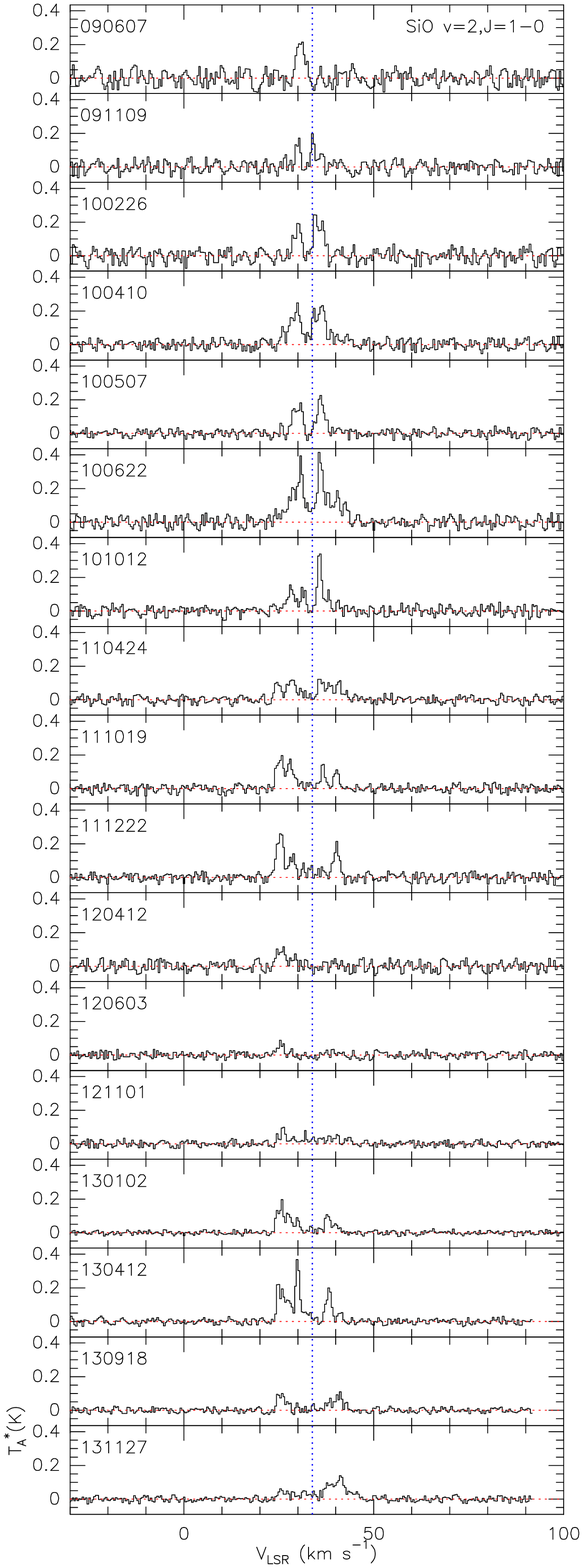}
   \vspace{-5mm}
   \caption{spectra of the SiO $v$ = 2, $J$ = 1$\rightarrow$0 maser monitored toward OH231.8.}
   \label{fig:b3}
\end{figure}

\begin{figure}
   \includegraphics[width=\columnwidth]{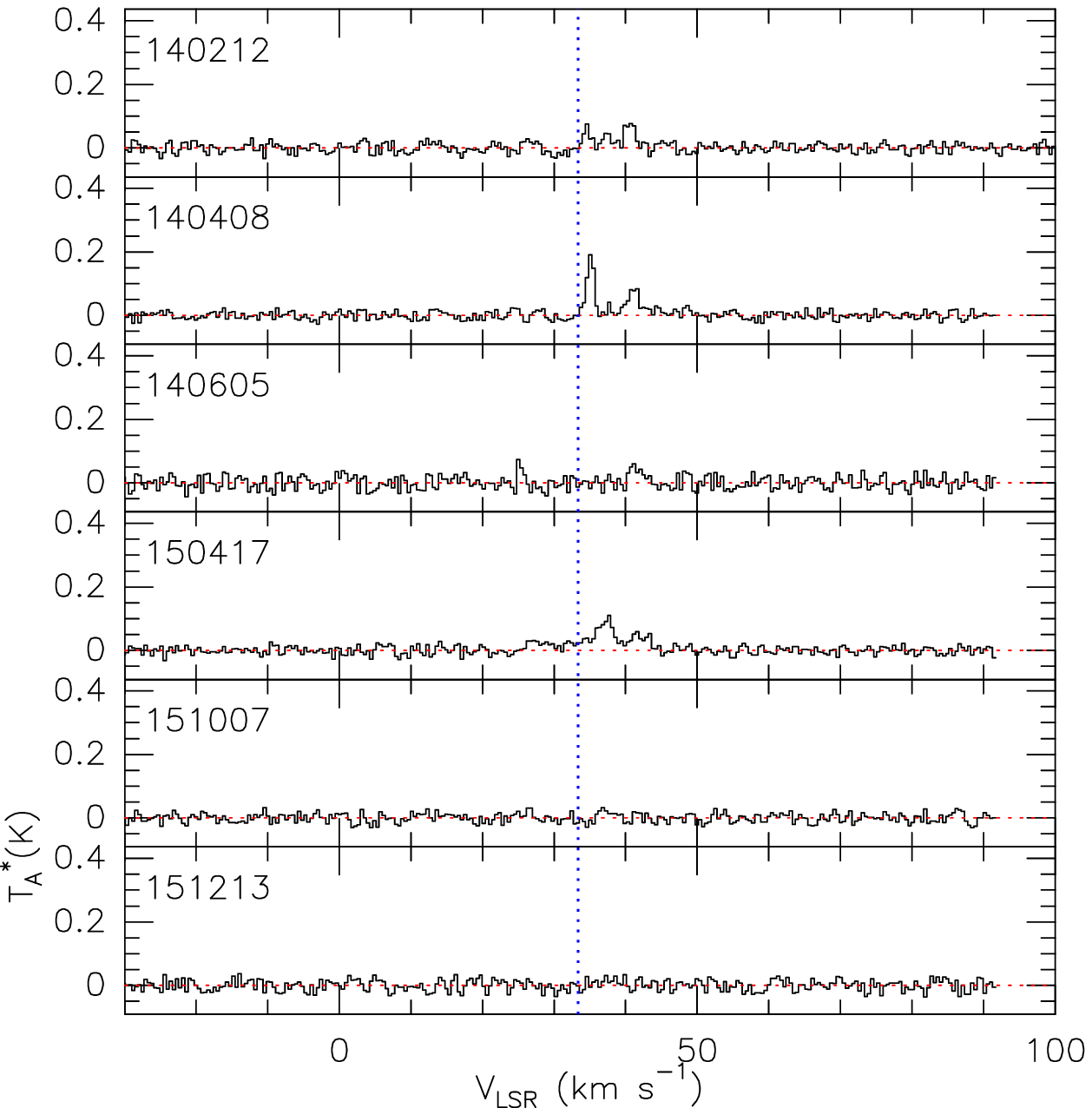}
   \vspace{-5mm}
   \contcaption{.}
   \label{fig:continued}
\end{figure}

\end{document}